\documentclass[aps,prb,twocolumn,superscriptaddress,dvips]{revtex4}
\usepackage{feynmp}
\usepackage{amsmath}
\usepackage{amssymb}
\usepackage{epsfig}
\usepackage{graphicx}
\usepackage{subfigure}
\usepackage{hyperref}
\usepackage{bbm}
\usepackage{color}

\newcommand{\Rmnum}[1]{\expandafter\@slowromancap\romannumeral #1@}
\allowdisplaybreaks[3]

\begin{document}

\title{Towards exact solutions for the superconducting $T_c$ induced by electron-phonon
interaction}

\author{Guo-Zhu Liu}
\altaffiliation{Corresponding author: gzliu@ustc.edu.cn}
\affiliation{Department of Modern Physics, University of Science and
Technology of China, Hefei, Anhui 230026, P. R. China}
\author{Zhao-Kun Yang}
\affiliation{Department of Modern Physics, University of Science and
Technology of China, Hefei, Anhui 230026, P. R. China}
\author{Xiao-Yin Pan}
\affiliation{Department of Physics, Ningbo University, Ningbo,
Zhejiang 315211, P. R. China}
\author{Jing-Rong Wang}
\affiliation{Anhui Province Key Laboratory of Condensed Matter
Physics at Extreme Conditions, High Magnetic Field Laboratory of the
Chinese Academy of Sciences, Hefei, Anhui 230031, P. R. China}

\begin{abstract}
Electron-phonon interaction plays an important role in metals and
can lead to superconductivity and other instabilities. Previous
theoretical studies on superconductivity are largely based on the
Migdal-Eliashberg theory, which neglects all the vertex corrections
to electron-phonon coupling and breaks down in many unconventional
superconductors. Here, we go beyond the Migdal-Eliashberg
approximation and develop a nonperturbative Dyson-Schwinger equation
approach to deal with the superconducting transition. Remarkably, we
take into account all the vertex corrections by solving two coupled
Ward-Takahashi identities derived from two global U(1) symmetries
and rigorously prove that the fully renormalized electron propagator
satisfies a self-closed integral equation that is directly amenable
to numerical computations. Our approach works equally well in the
weak and strong coupling regimes and provides an efficient method to
determine superconducting $T_c$ and other quantities. As an
application, our approach is used to investigate the high-$T_c$
superconductivity in one-unit-cell FeSe/SrTiO$_3$.
\end{abstract}

\maketitle


\section{Introduction}\label{sec:introduction}

In crystalline solids, atoms are arranged in a highly ordered
pattern, forming periodic lattices. Phonons, the quanta of
collective vibrating modes of atoms, interact with the mobile
electrons of metals. The electron-phonon interaction (EPI) plays a
major role in all metals \cite{Schrieffer64, AGD, Scalapino}, and
governs many thermodynamic and transport properties. Under proper
conditions, EPI may induce a number of possible phase-transition
instabilities, such as superconductivity and charge density wave
(CDW). Metals cannot be thoroughly understood without detailed
knowledge of EPI. Finding a reliable method to efficiently treat the
EPI-induced quantum many-body effects is one of the greatest
challenges in condensed-matter physics.

\begin{figure}[htbp]
\centering
\includegraphics[width=3.0in]{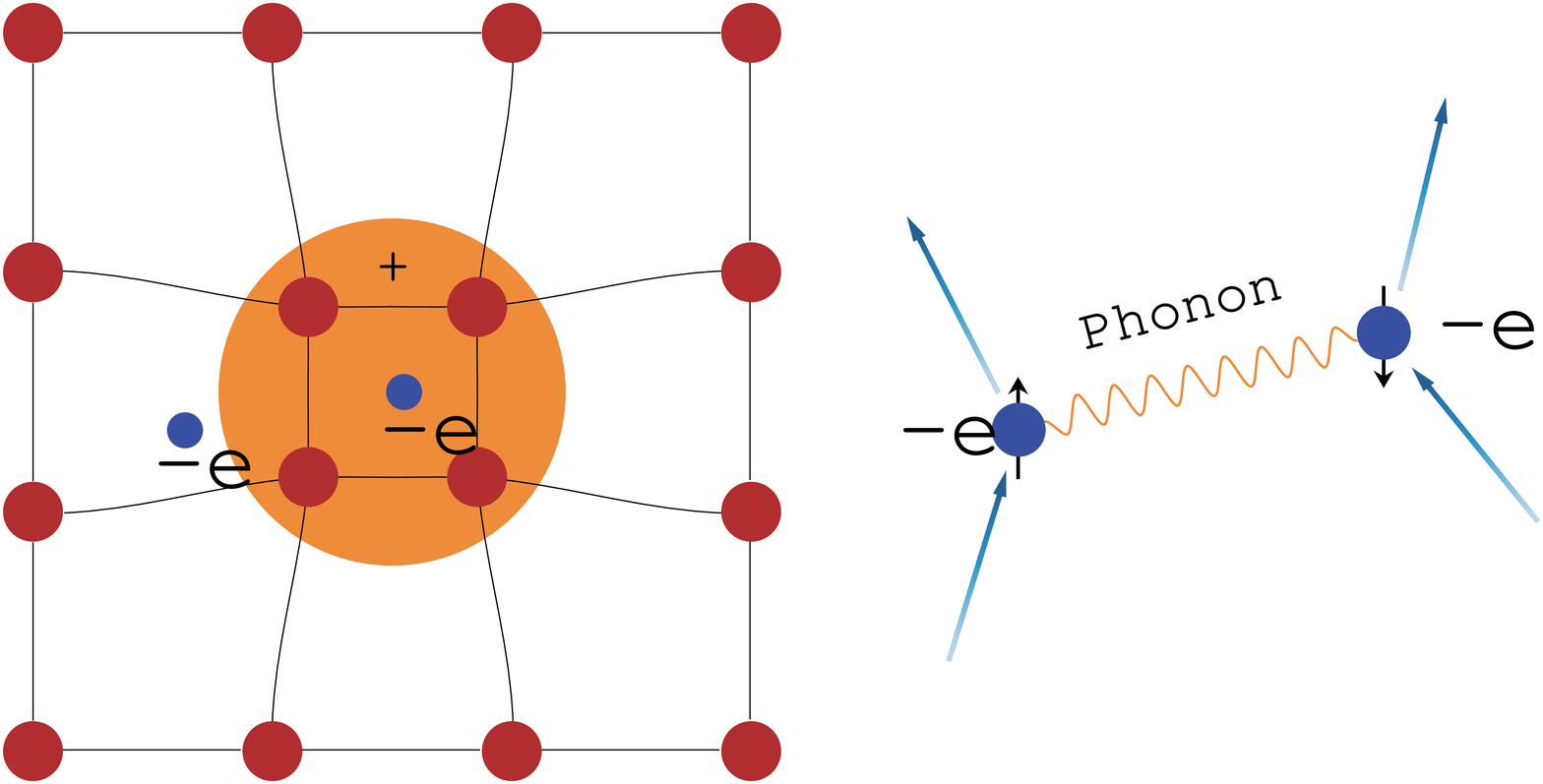}
\caption{Schematic illustration of electron-phonon interaction. A
mobile electron attracts another mobile electron by exchanging
phonons. Two electrons with opposite momenta and spin directions are
combined to form a Cooper pair, which is a composite boson.
Superconductivity is realized as Cooper pairs condensate at low
temperatures.}\label{fig:interaction}
\end{figure}

A remarkable consequence of EPI is the realization of
superconductivity. Comparing to CDW and other phases,
superconductivity is more universal, of broader interest, and also
has much more technical and industrial applications. We believe that
superconductivity provides an ideal framework to develop new
nonperturbative quantum many-body methods. According to
Bardeen-Cooper-Schrieffer (BCS) theory \cite{Schrieffer64}, a
sufficiently strong EPI triggers Cooper pairing (see
Fig.~\ref{fig:interaction} for a schematic illustration) and then
leads to superconductivity. It has been established that EPI is
responsible for the onset of superconductivity in a large number of
conventional \cite{Schrieffer64, AGD, Scalapino, Migdal, Eliashberg,
Allen, Carbotte, Marsiglio19} and unconventional \cite{Xue12,
Shen14, Lee15, Gorkov16, Johnston-NJP2016, Martin19}
superconductors. To understand the properties of these
superconductors, it is important to find a quantitatively reliable
tool to accurately compute the superconducting transition
temperature $T_c$ and other relevant quantities. Without such a
tool, it would be hard to predict and apply realistic
superconducting materials. While the BCS theory identifies the
correct microscopic mechanism of superconductivity, it is a
mean-field theory and cannot compute the accurate value of $T_c$ in
most superconductors. The oversimplified BCS theory can be improved
by the Migdal-Eliashberg (ME) theory \cite{Migdal, Eliashberg},
which incorporates the retardation of phonon propagation, electron
mass renormalization, and Cooper pairing in a self-consistent
manner.

In the past sixty years, the ME theory has been extensively adopted
to investigate EPI-induced effects in numerous superconductors
\cite{Schrieffer64, AGD, Scalapino, Allen, Carbotte, Marsiglio19},
and is widely regarded as the standard theory of conventional
superconductivity. Specifically, it plays an overwhelmingly dominant
role \cite{Allen} in the computation of superconducting $T_c$. The
reliability of ME theory depends heavily on the validity of Migdal
theorem \cite{Migdal}, which states that all the quantum corrections
to the EPI vertex function $\Gamma_{\mathrm{v}}(q,p)$ are suppressed
by the small factor $\lambda(\omega_D/E_F)$, where $\lambda$ is a
dimensionless coupling constant, $\omega_D$ is the phonon frequency,
and $E_F$ is the Fermi energy, and therefore can be completely
ignored if $\lambda(\omega_D/E_F) \ll 1$. The ME results are
expected to be reliable as long as the ratio $\omega_D/E_F$ is
sufficiently small and/or the EPI is sufficiently weak. However, it
has long been recognized that the Migdal theorem is not always valid
\cite{Engelsberg63, Alexandrov01, Kivelson18}. There exist several
classes of superconductors in which $\lambda(\omega_D/E_F)$ is not
small. Notable examples include low carrier-density superconductors
such as SrTiO$_3$ \cite{Schooley64, Chubukov19} and Moir\'{e}
superconductor \cite{Caoyuan18, Martin19}, fulleride superconductors
\cite{Gunnarsson97, Cappelluti01}, cuprate superconductors
\cite{Carbotte, Shen05}, and one-unit-cell (1UC) FeSe/SrTiO$_3$
system \cite{Xue12, Shen14, Lee15, Gorkov16, Johnston-NJP2016}. The
ME results become especially unreliable when EPI gets strong. This
fact has already been discussed \cite{Engelsberg63, Alexandrov01,
Gunnarsson97, Cappelluti01} for decades, and recently was
re-confirmed by a determinant quantum Monte Carlo (DQMC) study
\cite{Kivelson18}. To describe EPI systems in which the Migdal
theorem breaks down, it is necessary to develop a more powerful
approach that can take into account all the potentially important
contributions omitted in the ME theory and meanwhile is valid in
both the weak and strong regimes of EPI.

Dyson-Schwinger (DS) equations refer to an infinite number of
self-consistently coupled integral equations of $n$-point ($n \geq
2$) correlation functions. All the interaction-induced effects are
embodied in these equations. The DS equation approach treats the
interacting electrons and phonons on an equal footing at the outset,
and is much more generic than ME formalism. Unfortunately, the full
set of DS equations are usually not closed. This seriously hinders
their applicability. To make the DS equations closed, one might
invoke a hard truncation (e.g., choosing some special Feynman
diagrams), or introduce an \emph{Ansatz} for the vertex function
$\Gamma_{\mathrm{v}}(q,p)$. However, such treatments are actually
based on unjustified assumptions and cannot be trusted.

In this paper, we will go beyond the traditional ME theory and
develop an efficient DS equation approach to accurately treat EPI.
In particular, we prove that the DS integral equation of the fully
renormalized electron propagator, denoted by $G(p)$, is indeed
self-closed. It is well known that, the main difficulty in acquiring
a more general theoretical description than the ME one is the
extreme complexity of the full vertex function
$\Gamma_{\mathrm{v}}(q,p)$. The full $\Gamma_{\mathrm{v}}(q,p)$
contains an infinite number of Feynman diagrams. Calculating all of
such diagrams seems to be a mission impossible. Actually, it is
already very difficult to compute the simplest one-loop diagram of
$\Gamma_{\mathrm{v}}(q,p)$, let alone those multi-loop diagrams.
Remarkably, in this paper, we find it possible to take all the EPI
vertex corrections into account without ignoring any Feynman
diagram. This is achieved by properly utilizing several
symmetry-imposed constraints on $2$- and $3$-point correlation
functions. In particular, we derive two coupled Ward-Takahashi
identities (WTIs) from the global U(1) symmetries of the system and
then incorporate all the vertex corrections after solving these two
WTIs. Based on these results, we can prove that the DS equation of
$G(p)$ is decoupled entirely from the DS equations of the full
phonon propagator $F(q)$ and other correlation functions. After
solving the self-closed integral equation of $G(p)$, one can
calculate the superconducting $T_c$ and other physical quantities
with high precision. Our approach is strictly nonperturbative and
does not involve any small expansion parameter. Thus this approach
is well applicable in the strong EPI regime.

As an application of our approach, we will investigate the
high-$T_c$ superconductivity induced by the interfacial optical
phonons (IOPs) in 1UC FeSe/SrTiO$_3$ system, and examine how the
value of $T_c$ is affected by EPI vertex corrections. It is found
that neglecting the vertex corrections may significantly
underestimate $T_c$. This result would help ascertain whether the
coupling of electrons in FeSe film to IOPs by itself is able to
produce the observed high $T_c$.

The rest of the paper is organized as follows. We define the
Lagrangian density of EPI systems in Sec.~\ref{sec:model}, and
present and analyze three coupled DS integral equations in
Sec.~\ref{sec:DS}. We make a detailed symmetry analysis and derive
two coupled WTIs in Sec.~\ref{sec:WTI}. We obtain an exact relation
between the EPI and current vertex functions within the framework of
functional integral in Sec.~\ref{sec:relation}. Using these results,
we derive the self-closed equations of mass renormalization function
and pairing function in Sec.~\ref{sec:equation}. We then apply the
DS equation approach to compute the superconducting $T_c$ based on a
simple model of 1UC FeSe/SrTiO$_3$ in Sec.~\ref{sec:application}. We
collect some basic rules of the functional integral in
Appendix~\ref{appendix:functional} and show how to derive the DS
equations of electron and phonon propagators in
Appendix~\ref{appendix:DS}. We present the linearized gap equations
in Appendix~\ref{appendix:system}, and demonstrate how the iterative
method works in Appendix~\ref{appendix:numerical}.

\section{The Model}\label{sec:model}

In order to deal with the formation of Cooper pairs, here we define
the standard Nambu spinor \cite{Nambu60}
\begin{eqnarray}
\Psi^{\dag}(\mathbf{p}) = (\psi_\uparrow^{\dag}(\mathbf{p}),
\psi_{\downarrow}(-\mathbf{p})).
\end{eqnarray}
The $d$-dimensional Lagrangian density \cite{Nambu60, Engelsberg63}
is
\begin{eqnarray}
\mathcal{L} &=& \Psi^{\dag}(p)(\epsilon \sigma_0
-\xi_{\mathbf{p}}\sigma_3)\Psi(p) + \frac{1}{2}
\phi^{\dag}(q)F_{0}^{-1}(q)\phi(q) \nonumber \\
&& - g\phi(q)\Psi^\dag(p+q)\sigma_3\Psi(p), \label{eq:Lagrangianepi}
\end{eqnarray}
where $\xi_{\mathbf{p}}$ is electron energy and $g$ is coupling
constant for EPI. Here, $\sigma_{1,2,3}$ are the standard Pauli
matrices, and $\sigma_{0}$ denotes the unit $2\times 2$ matrix.
Shorthand notations $p \equiv (\epsilon,\mathbf{p})$, $q \equiv
(\omega,\mathbf{q})$, and $z \equiv (t,\mathbf{z})$ will be used
throughout the paper. The phonon field $\phi(q)$ satisfies the
relation $\phi^{\dag}(\mathbf{q}) = \phi(-\mathbf{q})$. For
simplicity, we first consider the simplest case and assume that
$\xi_{\mathbf{p}} = \frac{\mathbf{p}^2}{2m}-\mu$ with $\mu$ being
the chemical potential. The free phonon propagator $F_{0}(q)$ can be
identified as the Fourier transformed expression of
$\mathbb{D}^{-1}$, where $\mathbb{D} = -(\partial_{t}^2 +
\Omega_{\mathbf{q}})$ is the dynamical operator for phonon field
$\phi$ that satisfies the equation $\mathbb{D}\phi(t,\mathbf{q}) =
0$ in the non-interacting limit. Both the electron dispersion
$\xi_{\mathbf{p}}$ and the phonon dispersion $\Omega_{\mathbf{q}}$
are strongly material dependent, and can be determined by carrying
out first-principle calculations. We emphasize that our approach is
independent of the concrete expressions of $\xi_{\mathbf{p}}$ and
$\Omega_{\mathbf{q}}$, and even independent of whether the scalar
field $\phi(q)$ represents phonon or other types of boson. Later we
will discuss how to generalize our results to realistic metals where
$\xi_{\mathbf{p}}$ exhibits a more complicated
$\mathbf{p}$-dependence than $\xi_{\mathbf{p}} =
\frac{\mathbf{p}^2}{2m}-\mu$.

To illustrate how our approach works, let us first define several
quantities. In quantum many-body theory, one studies various
$n$-point correlation functions
\begin{eqnarray}
\langle \mathcal{O}_{1}\mathcal{O}_{2}...\mathcal{O}_{n}\rangle,
\end{eqnarray}
where $\mathcal{O}$'s are Heisenberg operators and
$\langle...\rangle$ stands for the manipulation of taking the
expectation value (more precisely, the statistical average over all
the possible field configurations). The theoretical descriptions of
quantum many-body effects are based on three elementary quantities:
the full electron propagator $G(p)= -i\langle\Psi
\Psi^{\dag}\rangle$, the full phonon propagator $F(q)=-i\langle \phi
\phi^{\dag}\rangle$, and the full EPI vertex function
$\Gamma_{\mathrm{v}}(q,p)$ that is defined by the relation
\begin{eqnarray}
F(q)G(p+q)\Gamma_{\mathrm{v}}(q,p)G(p) = \langle \phi \Psi
\Psi^{\dag}\rangle.\label{eq:EPIvertex}
\end{eqnarray}
These three quantities embody the quantum corrections to the free
electron term, free phonon term, and EPI-coupling term of the
Lagrangian density (2), respectively. In the non-interacting limit,
$G(p)$ is reduced to the free electron propagator
\begin{eqnarray}
G_{0}(p)= \frac{1}{\epsilon \sigma_{0} -
\xi_{\mathbf{p}}\sigma_{3}},
\end{eqnarray}
$F(q)$ is reduced to the free phonon propagator
\begin{eqnarray}
F_{0}(q) = \frac{2\Omega_{\mathbf{q}}}{\omega^{2}
-\Omega_{\mathbf{q}}^{2}},
\end{eqnarray}
and the EPI vertex function is reduced to its bare expression
\begin{eqnarray}
\Gamma_{\mathrm{v}}(q,p) \to g \sigma_{3}.
\end{eqnarray}
One of the key challenges of quantum many-body theory is to
determine the full propagators $G(p)$ and $F(q)$ on the basis of
free propagators $G_{0}(p)$ and $F_{0}(q)$.

\section{Dyson-Schwinger equations}\label{sec:DS}

This section is devoted to the derivation of DS equations of $G(p)$,
$F(q)$, and $\Gamma_{\mathrm{v}}(q,p)$. The functional integral
formalism of quantum field theory \cite{Itzykson} will be adopted.
By using this formalism, the DS equations and WTIs can be derived in
a compact and elegant manner. To generate various $n$-point
correlation functions, we introduce three external sources
$\eta^{\dag}$, $\eta$, and $J$ for field operators $\Psi$,
$\Psi^{\dag}$, and $\phi$, respectively, and then write the total
Lagrangian density $\mathcal{L}_{\mathrm{T}}$ in $d$-dimensional
real-space as
\begin{eqnarray}
\mathcal{L}_{\mathrm{T}} &=& \Psi^{\dag}(t,\mathbf{z})
(i\partial_{t}\sigma_0 - \xi_{\mathbf{\partial}}\sigma_3)
\Psi(t,\mathbf{z}) + \frac{1}{2} \phi^{\dag}(t,\mathbf{z})
\mathbb{D}_{z}\phi(t,\mathbf{z})
\nonumber \\
&& -g\phi(t,\mathbf{z})\Psi^\dag(t,\mathbf{z}) \sigma_3
\Psi(t,\mathbf{z})+J\phi+\Psi^\dag\eta+\eta^\dag\Psi,
\end{eqnarray}
where
\begin{eqnarray}
\mathbb{D}_{z} = -\frac{\partial_{t}^2 +
\Omega_{\mathbf{\partial_{\mathbf{z}}}}^2}{2
\Omega_{\mathbf{\partial_{\mathbf{z}}}}}.
\end{eqnarray}
Here we use $\xi_{\mathbf{\partial}}$ and
$\Omega_{\mathbf{\partial_{\mathbf{z}}}}$ to denote the real-space
correspondence of $\xi_{\mathbf{p}}$ and $\Omega_{\mathbf{q}}$,
respectively. The partition function can be formally written as
\begin{eqnarray}
Z(\eta^{\dag},\eta,J) = \int \mathcal{D}\Psi \mathcal{D}\Psi^{\dag}
\mathcal{D}\phi e^{i\int\mathcal{L}_{\mathrm{T}}
(\Psi,\Psi^{\dag},\phi;\eta^{\dag},\eta,J)}.
\end{eqnarray}
According to the calculations presented in Appendix~\ref{appendix:DS},
$G(p)$, $F(q)$, and $\Gamma_{\mathrm{v}}(q,p)$ satisfy the following DS
integral equations
\begin{eqnarray}
G^{-1}(p) &=& G_0^{-1}(p) - ig\int_{q} \sigma_3 G(p+q)
F(q)\Gamma_{\mathrm{v}}(q,p), \label{eq:DS-e0} \\
F^{-1}(q) &=& F_{0}^{-1}(q) + i g\int_{p}\mathrm{Tr}\left[\sigma_3
G(p+q)\Gamma_{\mathrm{v}}(q,p)G(p)\right], \nonumber\\
\\
\Gamma_{\mathrm{v}}(q,p) &=& g\sigma_3 - \int_{p'}G(p'+q)
\Gamma_\mathrm{v}(q,p')G(p')K_{4}(p,p',q).\nonumber \\
\end{eqnarray}
The integration over $d$-dimensional energy-momenta is henceforth
abbreviated as $\int_{q} \equiv \frac{d^dq}{(2\pi)^d}$.
$K_{4}(p,p',q)$ is the kernel function defined via $4$-point
correlation function $GG K_4 GG = \langle \Psi \Psi^{\dag}\Psi
\Psi^{\dag}\rangle$. These three DS equations are formally exact and
contain all the quantum many-body effects caused by EPI. However,
they appear to be too complicated to tackle. It turns out that they
are not closed because $K_{4}(p,p',q)$ satisfies its own DS integral
equation that is in turn coupled to the DS equations of five-, six-,
and higher-points correlation functions. Indeed, there is an
infinite hierarchy of coupled DS equations, which can never be
really solved.

The conventional ME theory assumes, based on the Migdal theorem,
that $$\Gamma_{\mathrm{v}}(q,p) \to g\sigma_{3},$$ which amounts to
discarding all the $n$-point correlation functions with $n>2$. In
actual applications, one usually further assumes that $$F(q) \to
F_0(q).$$ Then there is only one single DS equation of $G(p)$, which
is numerically solvable. This is exactly how conventional ME theory
works. Taking $\Gamma_{\mathrm{v}}(q,p) = g\sigma_{3}$ is reliable
when the parameter $\lambda(\omega_D/E_F)$ is small enough. However,
as pointed out in Sec.~\ref{sec:introduction}, this condition is not
satisfied in many unconventional superconductors. To investigate
systems in which the ME theory becomes unreliable, one needs to
carefully include higher-order corrections to the vertex function
$\Gamma_{\mathrm{v}}(q,p)$ and also those to the phonon propagator
$F(q)$. This is absolutely difficult. Remember that
$\Gamma_{\mathrm{v}}(q,p)$ contains an infinite number of Feynman
diagrams. It is impossible to compute all the diagrams.

In the past several decades, numerous theorists have proposed
various methods to investigate the impact of vertex corrections on
the value of superconducting $T_c$ and other physical quantities. It
is fair to say that all previous attempts are unsuccessful.
Generically, previous studies incorporate the vertex corrections by
employing two sorts of strategies:

1) Compute a small number of special diagrams of
$\Gamma_{\mathrm{v}}(q,p)$. This strategy is apparently not
justified since the vertex function receives contributions from an
infinite number of Feynman diagrams. In fact, this method is
\emph{ad hoc} if one could not prove that the chosen diagrams are
overwhelmingly more important than the omitted ones.

2) Introduce some kind of \emph{ansatz} for
$\Gamma_{\mathrm{v}}(q,p)$. The main drawback of this method is that
there is no reliable guiding principle to ensure the validity of the
\emph{ansatz}.

In this paper, we will investigate the impact of EPI vertex
corrections by employing an entirely different method. We are not
intended to compute any Feynman diagram nor to introduce any
\emph{ansatz}. Motivated by previous studies on the nonperturbative
effects of quantum gauge theories \cite{Takahashi86, Kondo97, He01},
we will perform a generic quantum-field-theoretic analysis, and
manage to find out a number of intrinsic relations between different
correlation functions based on careful symmetry considerations. We
will demonstrate that the symmetry constraints are powerful enough
to allow for an unambiguous determination of the full vertex
function $\Gamma_{\mathrm{v}}(q,p)$.

\section{Ward-Takahashi identities}\label{sec:WTI}

The aim of this section is to derive two exact identities that
connect the full electron propagator $G(p)$ to a special current
vertex function.

Consider the following two global U(1) transformations
\begin{eqnarray}
\Psi &\to& e^{i\alpha\sigma_3}\Psi,\label{eq:csymmetry} \\
\Psi &\to& e^{i\alpha\sigma_0}\Psi. \label{eq:ssymmetry}
\end{eqnarray}
It is easy to check the action $S = \int\mathcal{L}(z)\equiv \int
d^{d}z\mathcal{L}(z)$ is invariant under these two transformations.
Noether's theorem dictates that the symmetry given by
Eq.~(\ref{eq:csymmetry}) induces a conserved current $j^{c}_{\mu}
\equiv (j^{c}_{t},\mathbf{j}^{c})$, whose time and spatial
components are given by \cite{Nambu60}
\begin{eqnarray}
j^{c}_{t}(z) &=& \Psi^{\dag}(z)\sigma_{3}\Psi(z), \\
\mathbf{j}^{c}(z) &=& \frac{i}{2m}[(\mathbf{\nabla}
\Psi^{\dag}(z))\sigma_{0}\Psi(z) -
\Psi^{\dag}(z)\sigma_{0}(\mathbf{\nabla}\Psi(z))].\nonumber \\
\label{eq:ccurrent}
\end{eqnarray}
In the absence of external sources, this current is conserved,
namely $$\partial_{\mu}j^{c}_{\mu}=0,$$ corresponding to the
conservation of electric charge. The symmetry
Eq.~(\ref{eq:ssymmetry}) generates another conserved current
$j^s_{\mu} \equiv (j^s_{t},\mathbf{j}^s)$, whose time and spatial
components are given by \cite{Nambu60}
\begin{eqnarray}
j^s_{t}(z) &=& \Psi^{\dag}(z)\sigma_{0}\Psi(z), \\
\mathbf{j}^s(z) &=& \frac{i}{2m}[(\mathbf{\nabla}
\Psi^{\dag}(z))\sigma_{3}\Psi(z) -
\Psi^{\dag}(z)\sigma_{3}(\mathbf{\nabla}\Psi(z))].\nonumber \\
\label{eq:scurrent}
\end{eqnarray}
In the absence of external sources, this current is also conserved,
namely $$\partial_{\mu}j^{s}_{\mu}=0,$$ corresponding to the
conservation of spin. The time components of charge and spin
currents, i.e., $j^{c}_{t}(z)=\Psi^{\dag}(z)\sigma_{3}\Psi(z)$ and
$j^{s}_{t}(z)=\Psi^{\dag}(z)\sigma_{0}\Psi(z)$, can be used to
define two current vertex functions $\Gamma_{t}$ and $\Gamma_{s}$ as
follows
\begin{eqnarray}
&&\langle\Psi^{\dag}(z)\sigma_{3}\Psi(z)
\Psi(z_1)\Psi^{\dag}(z_2)\rangle \nonumber \\
&=& -\int dz_3 dz_4 G(z_1-z_3)
\Gamma_{t}(z,z_3,z_4)G(z_4-z_2), \label{eq:Gammatmain}\\
&&\langle\Psi^{\dag}(z)\sigma_{0} \Psi(z) \Psi(z_1)
\Psi^{\dag}(z_2)\rangle \nonumber \\
&=& -\int dz_3 dz_4 G(z_1-z_3) \Gamma_{s}(z,z_3,z_4) G(z_4-z_2).
\label{eq:Gammasmain}
\end{eqnarray}
The minus sign appearing on the right-hand side (r.h.s.) comes from
$i^{2}$ (recall that $\langle \Psi\Psi^{\dag}\rangle = iG$). The
propagator $G(z_1,z_3)$ depends only on the difference $z_1-z_3$ if
the system is translationally invariant. The Fourier transformation
of $\Gamma_{t,s}(z,z_3,z_4)$ is defined \cite{Engelsberg63, Kondo97,
He01} as
\begin{eqnarray}
\Gamma_{t,s}(z,z_3,z_4) &\equiv& \Gamma_{t,s}(z_3-z,z-z_4) \nonumber
\\
&=& \int_{q,p} e^{-i(p+q)(z_{3}-z)} e^{-ip(z-z_4)}
\Gamma_{t,s}(q,p). \label{eq:gammatsfourier}\nonumber
\\
\end{eqnarray}

The current vertex functions $\Gamma_{t,s}$ are defined through
conserved currents and satisfy some WTIs together with the full
electron propagator. Remarkably, $\Gamma_{t}$ would be entirely
determined and purely expressed in terms of full electron propagator
if one could find a sufficient number of WTIs. Later we will show
that two coupled WTIs suffice to determine $\Gamma_{t,s}$ in our
case.

The partition function $Z$ integrates over all possible field
configurations. Therefore, an infinitesimal variation of spinor
$\Psi$ should leave $Z$ unchanged. When $\Psi$ undergoes a generic
transformation $\Psi\to e^{i\alpha \sigma_m}\Psi$ where $\sigma_{m}$
might be $\sigma_{3}$ or $\sigma_{0}$, the action
$S[\Psi,\Psi^{\dag},\phi;\eta^{\dag},\eta,J]$ satisfies the
following equation
\begin{eqnarray}
0 = \langle\frac{\delta S}{i\delta\alpha}\rangle &=&
\langle-\frac{\delta S}{\delta\Psi}\frac{\delta\Psi}{i\delta\alpha} +
\frac{\delta\Psi^\dag}{i\delta\alpha} \frac{\delta S}{\delta
\Psi^\dag}\rangle \nonumber \\
&=& -\langle\frac{\delta S}{\delta \Psi} \sigma_{m} \Psi\rangle -
\langle\Psi^\dag\sigma_m^\dag\frac{\delta S}{\delta
\Psi^\dag}\rangle.
\end{eqnarray}
Substituting $S[\Psi,\Psi^{\dag},\phi;\eta^{\dag},\eta,J]$ into this
equation leads to
\begin{widetext}
\begin{eqnarray}
&&\langle\left(i\partial_{t} \Psi^\dag(z)\right)\sigma_0\sigma_m
\Psi(z)\rangle +\langle\left(\xi_{\partial_{\mathbf{z}}}
\Psi^\dag(z)\right)\sigma_3\sigma_m\Psi(z)\rangle +
\langle\Psi^\dag(z)\sigma_m^\dag\sigma_0 (i\partial_{t}
\Psi(z))\rangle-\langle\Psi^\dag(z) \sigma_m^\dag\sigma_3
(\xi_{\partial_{\mathbf{z}}}\Psi(z))\rangle
\nonumber \\
&=& \langle g\phi(z)\Psi^\dag(z)\left(\sigma_m^\dag\sigma_3 -
\sigma_3\sigma_m\right)\Psi(z) - \Psi^\dag(z)\sigma_m^\dag\eta(z)
+\eta^\dag(z)\sigma_m\Psi(z)\rangle.\label{eq:STIcurrent}
\end{eqnarray}
The first term of right-hand side (r.h.s.) of this identity is
induced by the EPI. It vanishes if we choose $\sigma_m=\sigma_3$ and
$\sigma_m=\sigma_0$. We obtain
\begin{eqnarray}
\langle i\partial_{t}\left(\Psi^\dag(z)\sigma_{3}
\Psi(z)\right)\rangle + \langle\left(\xi_{\partial_{\mathbf{z}}}
\Psi^{\dag}(z)\right)\sigma_{0} \Psi(z) - \Psi^\dag(z) \sigma_{0}
\left(\xi_{\partial_{\mathbf{z}}}\Psi(z)\right)\rangle = \langle
-\Psi^\dag(z)\sigma_{3} \eta(z) +
\eta^\dag(z)\sigma_{3}\Psi(z)\rangle \label{eq:STIchargemain}
\end{eqnarray}
for $\sigma_{m}=\sigma_{3}$, and
\begin{eqnarray}
\langle i\partial_{t}\left(\Psi^\dag(z)\sigma_{0}
\Psi(z)\right)\rangle + \langle\left(\xi_{\partial_{\mathbf{z}}}
\Psi^{\dag}(z)\right)\sigma_{3} \Psi(z) - \Psi^\dag(z)
\sigma_{3}\left(\xi_{\partial_{\mathbf{z}}} \Psi(z)\right)\rangle =
\langle -\Psi^\dag(z)\sigma_{0} \eta(z) + \eta^\dag(z)
\sigma_{0}\Psi(z)\rangle \label{eq:STIspinmain}
\end{eqnarray}
for $\sigma_{m}=\sigma_{0}$. The l.h.s. of
Eq.~(\ref{eq:STIchargemain}) is the expectation value of the
divergence of charge current $j^{c}_{\mu}$, since
\begin{eqnarray}
\langle i\partial_{\mu}j^{c}_{\mu}(z)\rangle &=& \langle
i\partial_{t}j^{c}_{t}(z)\rangle + \langle
i\mathbf{\nabla}\cdot \mathbf{j}^{c}(z)\rangle \nonumber \\
&=& \langle i\partial_t\left(\Psi^{\dag}(z)\sigma_{3}\Psi(z)\right)
\rangle - \frac{1}{2m}\langle \mathbf{\nabla}\cdot
\left[\left(\mathbf{\nabla} \Psi^{\dag}(z)\right) \sigma_{0}
\Psi(z)-\Psi^{\dag}(z)\sigma_{0} \left(\mathbf{\nabla}
\Psi(z)\right)\right]\rangle \nonumber \\
&=& \langle i\partial_{t}\left(\Psi^{\dag}(z)
\sigma_{3}\Psi(z)\right) \rangle +
\langle\left(\xi_{\mathbf{\partial_z}}\Psi^{\dag}(z)\right)
\sigma_{0}\Psi(z) - \Psi^{\dag}(z)\sigma_{0}
\left(\xi_{\mathbf{\partial_z}}\Psi(z)\right)\rangle.
\end{eqnarray}
The l.h.s. of Eq.~(\ref{eq:STIspinmain}) is the expectation value of
the divergence of charge current $j^{s}_{\mu}$, since
\begin{eqnarray}
\langle i\partial_{\mu}j^s_{\mu}(z)\rangle &=& \langle
i\partial_{t}j^s_{t}(z) \rangle + \langle i\mathbf{\nabla}\cdot
\mathbf{j}^s(z)
\rangle \nonumber \\
&=& \langle i\partial_{t}(\Psi^{\dag}(z)\sigma_{0}\Psi(z))\rangle -
\frac{1}{2m} \langle \mathbf{\nabla}\cdot
\left[(\mathbf{\nabla}\Psi^{\dag}(z))\sigma_{3}\Psi(z)-\Psi^{\dag}(z)
\sigma_{3}(\mathbf{\nabla}\Psi(z))\right]
\rangle \nonumber \\
&=& \langle i\partial_{t}(\Psi^{\dag}(z)\sigma_{0}\Psi(z))\rangle
+\langle (\xi_{\mathbf{\partial_z}}\Psi^{\dag}(z))\sigma_{3}
\Psi(z)-\Psi^{\dag}(z)\sigma_{3}(\xi_{\mathbf{\partial_z}}
\Psi(z))\rangle.
\end{eqnarray}
Therefore, for charge and spin currents, we obtain the following two
identities:
\begin{eqnarray}
&&\langle i\partial_{\mu}j^{c}_{\mu}(z)\rangle = \langle
-\Psi^\dag(z)\sigma_{3} \eta(z) +
\eta^\dag(z)\sigma_{3}\Psi(z)\rangle, \label{eq:STIcharge1main} \\
&& \langle i\partial_{\mu}j^{s}_{\mu}(z)\rangle = \langle
-\Psi^\dag(z)\sigma_{0}\eta(z) +
\eta^\dag(z)\sigma_{0}\Psi(z)\rangle. \label{eq:STIspin1main}
\end{eqnarray}
These two identities are called Slavnov-Taylor identities (STIs).
STIs play a significant role in the proof of renormalization of
quantum gauge theories \cite{Itzykson}, and also are of paramount
importance in our analysis. One can regard STIs as the
generalization of Noether theorem to the case in which the system is
coupled to external sources. Once external sources are removed, the
above two STIs will be reduced to
$\langle\partial_{\mu}j^{c,s}_{\mu}\rangle=0$, which naturally
reproduces the Noether's theorem.

Our ultimate goal is to derive the relation between the current
vertex functions $\Gamma_{t,s}$ defined in
Eqs.~(\ref{eq:Gammatmain}-\ref{eq:Gammasmain}) and the full electron
propagator. This can be achieved by calculating functional
derivatives of STIs with respect to external sources. Performing
functional derivatives of Eq.~(\ref{eq:STIcharge1main}) and
Eq.~(\ref{eq:STIspin1main}) with respect to $\eta(z_2)$ and then to
$\eta^{\dag}(z_1)$ yields
\begin{eqnarray}
\langle i\partial_{\mu}j^{c}_{\mu}(z) \Psi(z_1)\Psi^{\dag}(z_2)
\rangle &=& -G(z_1-z)\sigma_3 \delta(z-z_2) \nonumber
\\
&& +\delta(z-z_1)\sigma_3 G(z-z_2), \label{eq:STIcharge2main} \\
\langle i\partial_{\mu}j^{s}_{\mu}(z)\Psi(z_1)\Psi^{\dag}(z_2)
\rangle &=& -G(z_1-z)\sigma_{0}\delta(z-z_2) \nonumber
\\
&& +\delta(z-z_1)\sigma_{0} G(z-z_2). \label{eq:STIspin2main}
\end{eqnarray}
The correlation function $\langle i\partial_{\mu}j^{c}_{\mu}(z)
\Psi(z_1)\Psi^\dag(z_2)\rangle$ can be divided into two parts,
namely
\begin{eqnarray}
\langle i\partial_{t}\left(\Psi^{\dag}(z)\sigma_{3}\Psi(z)\right)
\Psi(z_1)\Psi^\dag(z_2)\rangle
\end{eqnarray}
and
\begin{eqnarray}
\langle\left[\left(\xi_{\mathbf{\partial_z}}\Psi^{\dag}(z)\right)
\sigma_{0}\Psi(z) - \Psi^{\dag}(z)
\sigma_{0}\left(\xi_{\mathbf{\partial_z}}\Psi(z)\right)\right]
\Psi(z_1)\Psi^\dag(z_2)\rangle. \nonumber \\
\end{eqnarray}
For the first term, the time derivative $i\partial_{t}$ operates on
the product $\Psi^{\dag}(z)\sigma_{3}\Psi(z)$ as a whole and thus
can be directly moved out of the expectation value. Then it becomes
$i\partial_t \langle \Psi^{\dag}(z)\sigma_{3}\Psi(z)
\Psi(z_1)\Psi^\dag(z_2)\rangle$, which can be expressed via the
current vertex function $\Gamma_{t}$ defined in
Eq.~(\ref{eq:Gammatmain}). The second term needs to be treated
carefully. Since $\xi_{\mathbf{\partial_z}}$ operates solely on
$\Psi^{\dag}(z)$ or solely on $\Psi(z)$, it cannot be directly moved
out. Here, it is interesting to notice that, the second term is
formally very similar to the current vertex function $\Gamma_{s}$
defined by Eq.~(\ref{eq:Gammasmain}). Remember that $\Gamma_{s}$ is
defined by $\langle\Psi^{\dag}(z)\sigma_{0} \Psi(z)
\Psi(z_1)\Psi^{\dag}(z_2)\rangle$, which does not contain
differential operators inside expectation value. In order to relate
the second term to $\Gamma_{s}$, we need to move the operator
$\xi_{\partial_{\mathbf{z}}}$ out of the statistical average. This
purpose can be achieved by adopting the point-splitting technique.

The point-splitting technique was first proposed by Dirac
\cite{Dirac} and afterwards discussed by other theorists
\cite{Peierls, Serber}. In an influential paper \cite{Schwinger},
Schwinger used this technique to demonstrate how to maintain the
relativistic and gauge invariance of physical quantities in the
context of quantum electrodynamics. Subsequently, this technique was
employed by several theorists \cite{Jackiw69, Callan70, Bardeen69}
to analyze the anomalies of neutral axial-vector current. Nowadays
the point-splitting technique has already been developed into a
well-established method and widely used in high-energy physics
\cite{Peskin, Schnabl}. The basic idea of this technique is simple:
split the point at which the product of two spinor fields are
located into two slightly separated points and let the two points
coincide at the end of calculations.

In our case, we will split $z$ into two different points denoted by
$z$ and $z^{\prime}$, perform Fourier transformations, and finally
take the limit $z^{\prime} \to z$ after all calculations are
completed. The main problem with the point-splitting technique is
that it could break the Lorentz invariance and the local gauge
invariance when applied to quantum gauge theories and thus needs to
be implemented with great caution. However, this problem is not
encountered in our case because the EPI system under consideration
exhibits neither Lorentz invariance nor local gauge invariance. Our
analytical derivation is based on the global gauge invariance and
the translational invariance, which are not spoiled by
point-splitting manipulation.

Using the point-slitting technique \cite{He01}, we now re-cast
Eq.~(\ref{eq:STIcharge2main}) and Eq.~(\ref{eq:STIspin2main}) in the
following forms
\begin{eqnarray}
&&i\partial_{t}\langle\Psi^\dag(z)\sigma_3\Psi(z)
\Psi(z_1)\Psi^{\dag}(z_2)\rangle + \lim_{z^\prime\to z}
(\xi_{\mathbf{\partial_{z^\prime}}}-\xi_{\mathbf{\partial_z}})
\langle\Psi^\dag(z^\prime)\sigma_0\Psi(z)\Psi(z_1)
\Psi^{\dag}(z_2)\rangle \nonumber \\
&=& -G(z_1-z)\sigma_3 \delta(z-z_2)+\delta(z-z_1)\sigma_3 G(z-z_2),
\label{eq:STIcharge3main}
\end{eqnarray}
\begin{eqnarray}
&&i\partial_{t}\langle\Psi^\dag(z)\sigma_0\Psi(z)
\Psi(z_1)\Psi^{\dag}(z_2)\rangle + \lim_{z^\prime\to z}
(\xi_{\mathbf{\partial_{z^\prime}}}-\xi_{\mathbf{\partial_z}})
\langle\Psi^\dag(z^\prime)\sigma_3\Psi(z)\Psi(z_1)
\Psi^{\dag}(z_2)\rangle \nonumber \\
&=& -G(z_1-z)\sigma_{0}\delta(z-z_2)+\delta(z-z_1)\sigma_{0}
G(z-z_2). \label{eq:STIspin3main}
\end{eqnarray}
Making use of Eqs.~(\ref{eq:Gammatmain}-\ref{eq:Gammasmain}), we
find it sufficient to express these two STIs in terms of two current
vertex functions $\Gamma_{t}$ and $\Gamma_{s}$. Making Fourier
transformations to the above two STIs eventually leads to two WTIs:
\begin{eqnarray}
\omega\Gamma_{t}(q,p)-(\xi_{\mathbf{p}+\mathbf{q}} -
\xi_{\mathbf{p}})\Gamma_{s}(q,p) &=& G^{-1}(p+q)\sigma_{3} -
\sigma_{3}G^{-1}(p), \label{eq:WTI1main} \\
\omega\Gamma_{s}(q,p)-(\xi_{\mathbf{p}+\mathbf{q}}-\xi_{\mathbf{p}})
\Gamma_{t}(q,p) &=& G^{-1}(p+q)\sigma_{0} - \sigma_{0}G^{-1}(p).
\label{eq:WTI2main}
\end{eqnarray}
\end{widetext}
Originally, charge conservation and spin conservation are
independent. They are expected to yield two independent WTIs that
contain four independent current vertex functions, two for
charge-related WTI and two for spin-related WTI. However, with the
help of point-splitting technique, we find that these four functions
are indeed related, and can be described by two independent
functions, namely $\Gamma_{t}(q,p)$ and $\Gamma_{s}(q,p)$.

In Ref.~\cite{Engelsberg63}, Engelsberg and Schrieffer have derived
the WTI induced by charge conservation. Using the Nambu spinor, that
WTI can be written in the form
\begin{eqnarray}
\omega\Gamma_{t}(q,p)-\mathbf{q}\cdot \mathbf{\Gamma}(q,p) =
G^{-1}(p+q)\sigma_{3} - \sigma_{3}G^{-1}(p).\label{eq:WTI-1963}
\end{eqnarray}
In their work, the vector function $\mathbf{\Gamma}(q,p)$ is
entirely unknown. This implies that the function $\Gamma_{t}(q,p)$
cannot be uniquely determined. For this reason, although the above
WTI has been known for nearly sixty years, it is of little use in
practical studies on EPI-induced effects. Comparing to
Ref.~\cite{Engelsberg63}, in this work we have obtained two
important new results. First, we have shown, making use of
point-splitting technique, that $\mathbf{\Gamma}(q,p)$ can be
expressed in terms of current vertex function $\Gamma_{s}(q,p)$ as
$$\mathbf{q}\cdot \mathbf{\Gamma}(q,p) = \left(\xi_{\mathbf{p+q}} -
\xi_{\mathbf{p}}\right)\Gamma_{s}(q,p).$$ Second, we have shown that
the WTI related to spin conservation can also be expressed in terms
of the two current vertex functions $\Gamma_{t}(q,p)$ and
$\Gamma_{s}(q,p)$. Since the two unknown functions $\Gamma_{t}(q,p)$
and $\Gamma_{s}(q,p)$ satisfy two coupled WTIs, they can be
unambiguously determined. By solving
Eqs.~(\ref{eq:WTI1main}-\ref{eq:WTI2main}), it is straightforward to
obtain
\begin{eqnarray}
\Gamma_{t}(q,p) &=& \frac{\omega[G^{-1}(p+q)\sigma_3 -\sigma_3
G^{-1}(p)]}{\omega^{2} -
(\xi_{\mathbf{p}+\mathbf{q}}-\xi_{\mathbf{p}})^2} \nonumber
\\
&& +\frac{(\xi_{\mathbf{p}+\mathbf{q}}-\xi_{\mathbf{p}})[G^{-1}(p+q)
\sigma_0-\sigma_0G^{-1}(p)]}{\omega^{2} -
(\xi_{\mathbf{p}+\mathbf{q}}-\xi_{\mathbf{p}})^2}.\nonumber \\
\label{eq:Gamma_tG}
\end{eqnarray}
The other function $\Gamma_{s}(q,p)$ can be easily derived, but it
is not directly useful at this stage and thus will not be given
explicitly.

It is now necessary to discuss the specific expression of fermion
dispersion $\xi_{\mathbf{p}}$. Remember our starting point is the
Lagrangian density (2). The Laplace operator has the standard form
$\xi_{\mathbf{\partial}} = -\frac{\nabla_{\mathbf{z}}^{2}}{2m}$,
since the motion of free electrons is supposed to satisfy
non-relativistic Schrodinger equation. It becomes $\xi_{\mathbf{p}}
= \frac{\mathbf{p}^{2}}{2m}-\mu$ after making Fourier transformation
and introducing chemical potential. Such an electron dispersion is
often oversimplified. In more realistic studies on metals, the
electron dispersion $\xi_{\mathbf{p}}$ is usually derived from a
certain lattice (tight-binding) model and exhibits a much more
complicated dependence on momenta. In this case, one should replace
the simple Laplace operator $-\frac{\nabla_{\mathbf{z}}^{2}}{2m}$
with a more generic operator $\xi_{\mathbf{\partial}}$, which can be
obtained from the dispersion $\xi_{\mathbf{p}}$ by doing Fourier
transformation. Normally, the generic operator
$\xi_{\mathbf{\partial}}$ could be expanded as the sum of various
powers of gradient operator $\mathbf{\nabla}$. Accordingly, the
operator $\mathbf{\nabla}$ appearing in the spatial components of
charge current $j^{c}_{\mu}$ (see Eq.~(\ref{eq:ccurrent})) and that
of spin current $j^{s}_{\mu}$ (see Eq.~(\ref{eq:scurrent})) should
be replaced by
$\xi_{\mathbf{\partial}}\frac{\mathbf{\nabla}}{\nabla^{2}}$. After
performing a series of calculations, one would still obtain the same
WTIs given by Eqs.~(\ref{eq:WTI1main}-\ref{eq:WTI2main}). Therefore,
these two WTIs are independent of the expressions of
$\xi_{\mathbf{p}}$.

\section{Relation between $\Gamma_{\mathrm{v}}(q,p)$ and
$\Gamma_t(q,p)$}\label{sec:relation}

Thus far we have defined and analyzed two sorts of vertex functions.
One is the EPI vertex function $\Gamma_{\mathrm{v}}(q,p)$ that is
defined through the mean value $\langle\phi\Psi\Psi^{\dag}\rangle$,
as shown in Eq.~(\ref{eq:EPIvertex}). $\Gamma_{\mathrm{v}}(q,p)$ is
a scalar function and enters into the DS equations of electron and
phonon propagators. Notice that $\Gamma_{\mathrm{v}}(q,p)$ itself
does not necessarily satisfy any WTI unless the boson field couples
to certain current operator composed of fermion fields. The other
sort is called the current vertex function, including
$\Gamma_{t}(q,p)$ and $\Gamma_{s}(q,p)$. Current vertex functions
are defined in terms of conserved currents and thus satisfy a number
of WTIs as the result of current conservation. These two sorts of
vertex functions are closely related but are apparently not
identical. Below we derive their relation.

In the last section, we have derived two WTIs based on the fact that
the partition function $Z$ is not changed by infinitesimal
variations of spinor field $\Psi$. Here, we require that $Z$ is
invariant under an infinitesimal variation of phonon field $\phi$.
This fact is described by the equation
$$0 = \int \mathcal{D}\phi\mathcal{D}\Psi^{\dag}\mathcal{D}\Psi
\frac{\delta}{\delta\phi}\exp\{iS\},$$ which then gives rise to
\begin{eqnarray}
0 &=& \langle \mathbb{D}_{\mathbf{z}}\phi(z) - g \Psi^\dag(z)
\sigma_3 \Psi(z) +J(z) \rangle,\label{eq:eomphi}
\end{eqnarray}
The above formula is re-written in the form
\begin{eqnarray}
g \langle\Psi^\dag(z) \sigma_3 \Psi(z)\rangle &=&
\mathbb{D}_{\mathbf{z}}\frac{\delta W}{\delta J(z)} + J(z),
\end{eqnarray}
which, after taking the functional derivative with respect to
$\eta^{\dag}$ and $\eta$ in order, leads to
\begin{eqnarray}
&&\frac{\delta^2}{\delta\eta^\dag(z_1)\delta\eta(z_2)}
\langle\Psi^\dag(z)\sigma_3 \Psi(z)\rangle \nonumber \\
&=& \langle \Psi^\dag(z)\sigma_3 \Psi(z) \Psi(z_1)
\Psi^{\dag}(z_2)\rangle \nonumber \\
&=& \langle j^{c}_t(z)\Psi(z_1)\Psi^\dag(z_2)\rangle \nonumber \\
&=& g^{-1} \mathbb{D}_{\mathbf{z}} \frac{\delta^3W}{\delta
J(z)\delta\eta^\dag(z_1)\delta\eta(z_2)}.\label{eq:eomphonon2}
\end{eqnarray}
Here, we have utilized an important fact that the electron density
operator $\Psi^\dag\sigma_3 \Psi$ that couples to phonon field
$\phi$ is proportional to the time-component of conserved charge
current $j^{c}_{\mu}$, i.e., $\Psi^\dag\sigma_3 \Psi = j^{c}_{t}$.

Making use of the identity
\begin{eqnarray}
\frac{\delta^3W}{\delta J\delta\eta^\dag\delta\eta} = - F G
\frac{\delta^3\Xi}{\delta\phi \delta\Psi^\dag \delta\Psi}G,
\end{eqnarray}
we obtain from Eq.~(\ref{eq:eomphonon2}) and
Eq.~(\ref{eq:Gammatmain}) that
\begin{eqnarray}
&&\int dz_3dz_4 G(z_1-z_3)\Gamma_{t}(z,z_3,z_4)G(z_4-z_2)
\nonumber \\
&=& \int dz_5 dz_3 dz_4 g^{-1} \mathbb{D} F(z,z_5) \nonumber
\\
&& \times G(z_1-z_3)\Gamma_{\mathrm{v}}(z_5,z_3,z_4)
G(z_4-z_2).\label{eq:gammatf0fggammavg}
\end{eqnarray}

After performing Fourier transformation,
Eq.~(\ref{eq:gammatf0fggammavg}) is transformed into an exact
identity
\begin{eqnarray}
\Gamma_{t}(q,p) = g^{-1}F_0^{-1}(q)F(q)\Gamma_{\mathrm{v}}(q,p),
\label{eq:GammaD0DGamma1}
\end{eqnarray}
which builds a connection among $F_{0}(q)$, $F(q)$, $\Gamma_t(q,p)$,
and $\Gamma_{\mathrm{v}}(q,p)$. This identity can be further written
as
\begin{eqnarray}
g F_{0}(q)\Gamma_{t}(q,p) = F(q) \Gamma_{\mathrm{v}}(q,p),
\label{eq:GammaD0DGamma}
\end{eqnarray}
Note that this identity was first obtained by Engelsberg and
Schrieffer \cite{Engelsberg63}. However, it turns out that they did
not realize the importance of this identity. As can be seen from
Appendix~\ref{appendix:DS} of Ref.~\cite{Engelsberg63}, after
obtaining the identity, they did not discuss its physical
implication but were trying to prove that $\Gamma_{\mathrm{v}}(q,p)$
satisfies the charge-related WTI. Nevertheless, it is
$\Gamma_{t}(q,p)$, rather than $\Gamma_{\mathrm{v}}(q,p)$, that
satisfies the WTI. To solve this problem, they took the
zero-momentum limit $\mathbf{q} \to 0$ and then argued that
$\Gamma_{t}(q,p) = \Gamma_{\mathrm{v}}(q,p)$ as $\mathbf{q} \to 0$.
Thus the EPI vertex $\Gamma_{\mathrm{v}}(q,p)$ approximately
satisfies the charge-related WTI in the special limit $\mathbf{q}
\to 0$. While their argument was absolutely correct, the potential
importance of the above identity was entirely overlooked.

In this paper, we have re-derived the identity given by
Eq.~(\ref{eq:GammaD0DGamma}) within the framework of functional
integral. The crucial new insight provided by our paper is that we
fully realize the importance of this identity and make use of its
general expression (without taking any limit) to prove that the DS
equation of the electron propagator $G(p)$ is decoupled from all the
other DS equations. This will be illustrated in the next section.

\section{Self-closed integral equation of electron propagator}
\label{sec:equation}

We now demonstrate how to use the identity of
Eq.~(\ref{eq:GammaD0DGamma}) to simplify DS equations. Originally,
the DS integral equations of $G(p)$ and $F(q)$ are coupled to each
other self-consistently, reflecting the dramatic mutual influence
between electrons and phonons. Their equations are further coupled
to the DS equations of all the $n$-point correlation functions with
$n>2$. It would be extremely difficult to solve such an infinite
number of coupled equations. To simplify these equations, we observe
that the product $F(q)\Gamma_{\mathrm{v}}(q,p)$ enters into the DS
equation of $G(p)$ as a whole. After inserting the identity
Eq.~(\ref{eq:GammaD0DGamma}) into Eq.~(\ref{eq:DS-e0}), we obtain
\begin{eqnarray}
G^{-1}(p) = G_0^{-1}(p) - ig^{2}\int_{q}\sigma_3 G(p+q)
F_{0}(q)\Gamma_{t}(q,p).\label{eq:DS}
\end{eqnarray}
Now, the DS equation of $G(p)$ contains only $G_{0}(p)$, $G(p)$,
$F_{0}(q)$, and $\Gamma_{t}(q,p)$, and therefore is decoupled
completely from that of $F(q)$. This is a vast simplification. We
have already shown in the last section that $\Gamma_{t}(q,p)$ is
solely determined by $G(q+p)$ and $G(p)$. Based on these results, we
have proved rigorously that the DS equation of $G(p)$ is
self-closed.

The full electron propagator $G(p)$ can be numerically computed once
the free propagators $G_{0}(p)$ and $F_{0}(q)$ are known.
Generically, one can expand \cite{Nambu60} $G(p)$ as follows
\begin{eqnarray}
G(\epsilon,\mathbf{p}) =
\frac{1}{A_{1}(\epsilon,\mathbf{p})\epsilon\sigma_{0} -
A_{2}(\epsilon,\mathbf{p})\xi_{\mathbf{p}}\sigma_{3} +
\Delta(\epsilon,\mathbf{p})\sigma_{1}},\label{eq:Full-e}
\end{eqnarray}
where $A_{1}(\epsilon,\mathbf{p})$ is the mass renormalization
function, $A_{2}(\epsilon,\mathbf{p})$ is the chemical potential
renormalization, and $\Delta(\epsilon,\mathbf{p})$ is the
superconducting pairing function. The true superconducting order
parameter is determined by the ratio
$\Delta(\epsilon,\mathbf{p})/A_{1}(\epsilon,\mathbf{p})$. As
demonstrated by Nambu \cite{Nambu60}, it is only necessary to
include the $\sigma_{1}$ term since the $\sigma_{2}$ term can be
easily obtained from the $\sigma_{1}$ term upon a simple
transformation. Substituting Eq.~(\ref{eq:Full-e}) into
Eq.~(\ref{eq:Gamma_tG}) and Eq.~(\ref{eq:DS}) would decompose the DS
equation of $G(p)$ into three self-consistent equations for
$A_{1,2}(\epsilon,\mathbf{p})$ and $\Delta(\epsilon,\mathbf{p})$,
which are amenable to numerical studies. The pairing function
$\Delta(\epsilon,\mathbf{p})$ is finite in the superconducting state
and decreases with growing $T$. The superconducting $T_c$ is just
the temperature at which $\Delta(\epsilon,\mathbf{p})$ goes to zero
from nonzero values.

There is a subtle issue here. The two coupled WTIs are derived from
two global U(1) symmetries. These two symmetries are both respected
in the normal phase. The superconducting phase is a little more
complicated. While superconducting pairing, described by a nonzero
$\Delta(\epsilon,\mathbf{p})$, preserves the U(1) symmetry related
to spin conservation, it spontaneously breaks the U(1) symmetry for
charge conservation. The spin-related WTI should be always correct,
but the charge-related WTI needs to be treated more carefully. Nambu
\cite{Nambu60} and Schrieffer \cite{Schrieffer64} addressed this
issue and assumed that the charge-related WTI has the same
expression in both the normal and superconducting phases. Based on
such an assumption, Nambu \cite{Nambu60} and Schrieffer
\cite{Schrieffer64} further demonstrated that the gauge invariance
of electromagnetic response functions is maintained even in the
superconducting phase. If this assumption was reliable, one could
substitute the general electron propagator given by
Eq.~(\ref{eq:Full-e}) into Eq.~(\ref{eq:Gamma_tG}) and express
$\Gamma_{t}(q,p)$ in terms of $A_{1}(\epsilon,\mathbf{p})$,
$A_{2}(\epsilon,\mathbf{p})$, and $\Delta(\epsilon,\mathbf{p})$.
However, it seems premature to accept the above assumption without
reservation. A complete understanding is still lacking. It is well
established that the Anderson-Higgs (AH) mechanism should be invoked
to accommodate U(1)-symmetry breaking \cite{Anderson63, Higgs64}
inside the superconducting phase. For that purpose, one needs to
promote the global U(1) symmetry, defined in
Eq.~(\ref{eq:csymmetry}) with a constant $\alpha$, to the local one
(defined by coordinate-dependent $\alpha(z)$), which is realized by
coupling some U(1) gauge boson to the electrons, such that the
massless Goldstone boson generated by U(1)-symmetry breaking can be
naturally eliminated. It is currently unclear how to reconcile the
AH mechanism with our DS equations approach. This is a highly
nontrivial issue that we leave for future research.

The absence of a complete understanding of the superconducting phase
might not be an obstacle if our interest is restricted to the
computation of superconducting $T_c$, because the pairing function
vanishes continuously as $T \to T_c$. In the vicinity of $T_c$, the
charge-related U(1) symmetry is still preserved and it is not
necessary to consider the AH mechanism. In practical applications of
our approach, one could drop the $\Delta$-dependence of
$\Gamma_{t}(q,p)$ and then insert it into the DS equation of $G(p)$.
For notational simplicity, here it is convenient to divide the
function $\Gamma_{t}(q,p)$ into two parts
\begin{eqnarray}
\Gamma_{t}(q,p) \equiv \Gamma_{t3}(q,p)\sigma_3 -
\Gamma_{t0}(q,p)\sigma_0,\label{eq:Gammattwocomponents}
\end{eqnarray}
where
\begin{eqnarray}
\Gamma_{t3}(q,p) &=& \frac{i\omega\left[A_1(p+q)(i\epsilon+i\omega)
-A_1(p)i\epsilon\right]}{(i\omega)^2 - (\xi_{\mathbf{p}+\mathbf{q}}
- \xi_{\mathbf{p}})^2} \nonumber \\
&& -\frac{(\xi_{\mathbf{p}+\mathbf{q}} - \xi_{\mathbf{p}})
\left[A_2(p+q)\xi_{\mathbf{p}+\mathbf{q}} -
A_2(p)\xi_{\mathbf{p}}\right]}{(i\omega)^2 -
(\xi_{\mathbf{p}+\mathbf{q}} - \xi_{\mathbf{p}})^2},
\nonumber \\
\Gamma_{t0}(q,p) &=& \frac{i\omega[A_2(p+q)
\xi_{\mathbf{p}+\mathbf{q}}-A_2(p)\xi_{\mathbf{p}}]}{(i\omega)^2 -
(\xi_{\mathbf{p}+\mathbf{q}}-\xi_{\mathbf{p}})^2} \nonumber \\
&& -\frac{(\xi_{\mathbf{p}+\mathbf{q}}-\xi_{\mathbf{p}})[A_1(p+q)
(i\epsilon+i\omega)-A_1(p)i\epsilon]}{(i\omega)^2 -
(\xi_{\mathbf{p}+\mathbf{q}}-\xi_{\mathbf{p}})^2}. \nonumber
\end{eqnarray}
With the help of $\Gamma_{t3}(q,p)$ and $\Gamma_{t0}(q,p)$, it is
easy to derive from Eq.~(\ref{eq:DS}) the following three integral
equations:
\begin{widetext}
\begin{eqnarray}
A_1(p)i\epsilon &=& i\epsilon + g^2 \int\frac{d^dq}{(2\pi)^d}
\frac{2\Omega_\mathbf{q}}{\omega^2 +
\Omega_{\mathbf{q}}^2}\frac{A_1(p+q)(i\epsilon+i\omega)
\Gamma_{t3}(q,p) - A_2(p+q)\xi_{\mathbf{p+q}}
\Gamma_{t0}(q,p)}{A_1^2(p+q)(\epsilon+\omega)^2 +
A_2^2(p+q)\xi_{\mathbf{p+q}}^2+\Delta^{2}(p+q)},
\label{eq:A1equation} \\
A_2(p)\xi_{\mathbf{p}} &=& \xi_{\mathbf{p}} + g^2 \int
\frac{d^dq}{(2\pi)^d}\frac{2\Omega_\mathbf{q}}{\omega^2 +
\Omega_{\mathbf{q}}^2}\frac{A_1(p+q)(i\epsilon+i\omega)
\Gamma_{t0}(q,p) - A_2(p+q)\xi_{\mathbf{p+q}}
\Gamma_{t3}(q,p)}{A_1^2(p+q)(\epsilon + \omega)^2 +
A_2^2(p+q)\xi_{\mathbf{p+q}}^2+\Delta^{2}(p+q)},
\label{eq:A2equation} \\
\Delta(p) &=& g^2 \int \frac{d^dq}{(2\pi)^d}
\frac{2\Omega_\mathbf{q}}{\omega^2+\Omega_{\mathbf{q}}^2}
\frac{\Delta(p+q)\Gamma_{t3}(q,p)}{A_1^2(p+q)(\epsilon+\omega)^2 +
A_2^2(p+q)\xi_{\mathbf{p+q}}^2+\Delta^{2}(p+q)}.
\label{eq:Deltaequation}
\end{eqnarray}
\end{widetext}
These integral equations are self-consistently coupled, which
describes the important fact that the mass renormalization, chemical
potential renormalization, and Cooper pairing can affect each other.
By numerically solving these equations at a series of different
temperatures, one will be able to obtain the superconducting $T_c$.
In addition, the detailed $p$-dependence of $A_{1}(p)$ and
$A_{2}(p)$ can also be simultaneously extracted from the numerical
solutions.

In the non-interacting limit, the fully renormalized electron
propagator $G(p)$ is reduced to the free one, namely $G(p) \to
G_{0}(p)$. Similarly, $F(q) \to F_{0}(q)$. Substituting $G_{0}(p)$
into Eq.~(\ref{eq:Gammattwocomponents}) leads to $\Gamma_{t}(q,p)\to
\sigma_{3}$. After substituting $G_{0}(p)$ and $\Gamma_{t}(q,p)=
\sigma_{3}$ into Eqs.~(\ref{eq:A1equation}-\ref{eq:Deltaequation}),
one obtains the well known ME equations \cite{Schrieffer64, AGD,
Scalapino, Eliashberg, Allen}. Thus, the conventional ME equations
are a special limit of the more generic DS equations given by
Eqs.~(\ref{eq:A1equation}-\ref{eq:Deltaequation}).

Once the equations of $A_{1,2}(\epsilon,\mathbf{p})$ and
$\Delta(\epsilon,\mathbf{p})$ are numerically solved, the solutions
can be substituted to the DS equation of the phonon propagator
$F(q)$. Using the previously derived identities, we obtain
\begin{eqnarray}
F(q) = F_{0}(q) + ig^{2}F_{0}^{2}(q) \int_{p}
\mathrm{Tr}\left[\sigma_3G(p+q)\Gamma_{t}(q,p)G(p)\right]. \nonumber
\\
\end{eqnarray}
Since $\Gamma_{t}(q,p)$ is expressed in terms of $G(p)$ and
$G(p+q)$, one can extract the full information about the phonons by
directly integrating over $p\equiv (\epsilon,\mathbf{p})$, which is
easier than solving self-consistent integral equations. The full
phonon self-energy, also known as polarization function, then can be
computed based on $G(p)$ as follows
\begin{eqnarray}
\Pi(q) &=& F_{0}^{-1}(q) - F^{-1}(q), \nonumber \\
&=& -i g\int_{p}\mathrm{Tr}\left[\sigma_3
G(p+q)\Gamma_{\mathrm{v}}(q,p)G(p)\right], \nonumber \\
&=& \frac{-i g^{2}\int_{p}\mathrm{Tr}\left[\sigma_3
G(p+q)\Gamma_{t}(q,p)G(p)\right]}{1-i g^{2}F_{0}(q)\int_{p}
\mathrm{Tr}[\sigma_3 G(p+q)\Gamma_t(q,p)G(p)]}.\nonumber
\end{eqnarray}
This expression would be very useful in the calculation of
density-density correlation function $\langle
\Psi^{\dag}\sigma_{3}\Psi \Psi^{\dag}\sigma_{3}\Psi\rangle$.
However, since the primary interest of this work is to study the
superconducting transition, we will not further discuss the
behaviors of $F(q)$ and $\Pi(q)$. Next, we focus on the self-closed
DS equation of $G(p)$ and use it to compute the superconducting
$T_c$.

The DS equations and the coupled WTIs are derived by carrying out
generic field-theoretical analysis. They are exact and
nonperturbative as long as the U(1) symmetries are preserved. No
small expansion parameter is employed in the derivation. This is
apparently distinct from traditional perturbation theories. As is
well known, the ME theory is developed based on series expansion in
powers of a small parameter $\lambda(\omega_D/E_F)$; it retains only
the leading order contribution and entirely discards all the rest
contributions. Such an approximation becomes invalid when
$\lambda(\omega_D/E_F)$ becomes large. In contrast, our DS equation
approach does not need any small parameter and does not discard any
Feynman diagram. This guarantees that the results extracted from the
solutions of DS equations are valid for any value of $\lambda$ and
any value of $\omega_D/E_F$, which is a significant advantage
compared to traditional ME theory. To what extend the results about
$G(p)$ and $F(q)$ are exact is mainly determined by the errors
generated in numerical integration, which can be gradually reduced
by costing reasonably more computer resources.

Another remarkable advantage of our approach is that, the inclusion
of vertex corrections does not increase the computational
difficulties. The generic DS equations and the simplified ME
equations can be solved by the same numerical skills. A standard
method of solving such equations is the iterative method, which
continues to use the old values of some unknown functions
($A_{1,2}(\epsilon,\mathbf{p})$ and $\Delta(\epsilon,\mathbf{p})$ in
our case) to generate new values until stable results of such
unknown functions are reached. We demonstrate how the iterative
method works in practice in Appendix~\ref{appendix:numerical}. The
computational time for obtaining stable results is mainly determined
by the dimensions of the multiple integral. In the practical
implementation of our approach, including vertex corrections only
alters the kernel function but does not change the dimensions of the
integral. Thanks to this crucial feature, the computational time
needed to solve DS equations (including all vertex corrections) is
not dramatically longer than that is needed to solve ME equations
(neglecting all vertex corrections). There is little difference
between the efficiency of solving the complete DS equations and that
of solving the simplified ME equations.

Besides diagrammatic techniques, strongly correlated electron
systems are often studied by means of several numerical methods,
among which DQMC simulation \cite{Assaad08} and dynamical mean field
theory (DMFT) \cite{DMFT} are the most frequently adopted. Different
from DQMC, our approach can directly access the ultra-low energy
regime (i.e., thermodynamic limit), and is not plagued by the
fermion-sign problem. While DMFT is exact only when the spatial
dimension $d-1$ is taken to the limit of infinity, our approach is
applicable to EPI systems defined in any spatial dimension,
including the most realistic cases of two and three dimensions.

\section{Application to small-$\mathbf{q}$ electron-phonon interaction}
\label{sec:application}

Our DS equation approach provides an efficient and universal tool to
study the superconducting transition mediated by EPI. To testify its
efficiency, we now apply it to a concrete example. Here, we choose
to study 1UC FeSe/SrTiO$_3$, where the origin of observed high-$T_c$
superconductivity has been intensively studied in recent years but
currently remains an open puzzle.

Bulk FeSe becomes a superconductor below $T_c \approx 8$K
\cite{Wu08}. When 1UC FeSe film is placed on SrTiO$_3$ substrate
\cite{Xue12}, its $T_c$ is dramatically promoted. This discovery has
opened a new route to engineering interfacial high-$T_c$
superconductors. An important issue is to find out the underlying
mechanism that gives rise to such a high $T_c$. It has been revealed
\cite{Rebec17} that, although charge carrier doping and
K-intercalation also enhance $T_c$, $T_c$ could be higher than $70$K
only when 1UC FeSe is at the interface to SrTiO$_3$ or other similar
substrates. Thus, the interfacial coupling must play a unique role.
Angle-resolved photoemission spectroscopy (ARPES) experiments have
provided strong evidence \cite{Shen14, Lee15} that the coupling of
electrons of FeSe-layer to IOPs generated by oxygen ions of
SrTiO$_3$ may account for the observed replica bands and high-$T_c$.

\begin{figure}[htbp]
\centering
\includegraphics[width=2.82in]{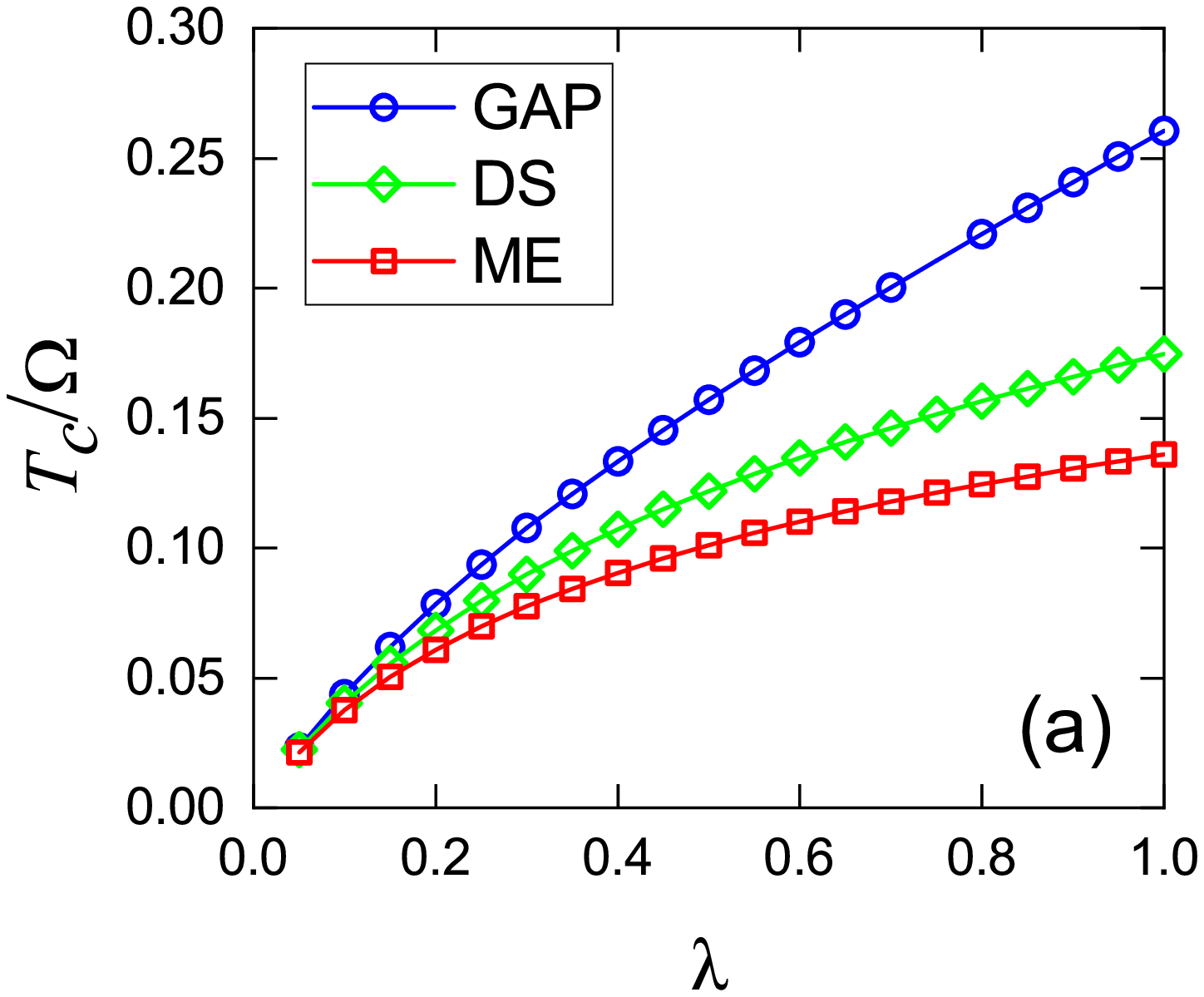}
\includegraphics[width=2.82in]{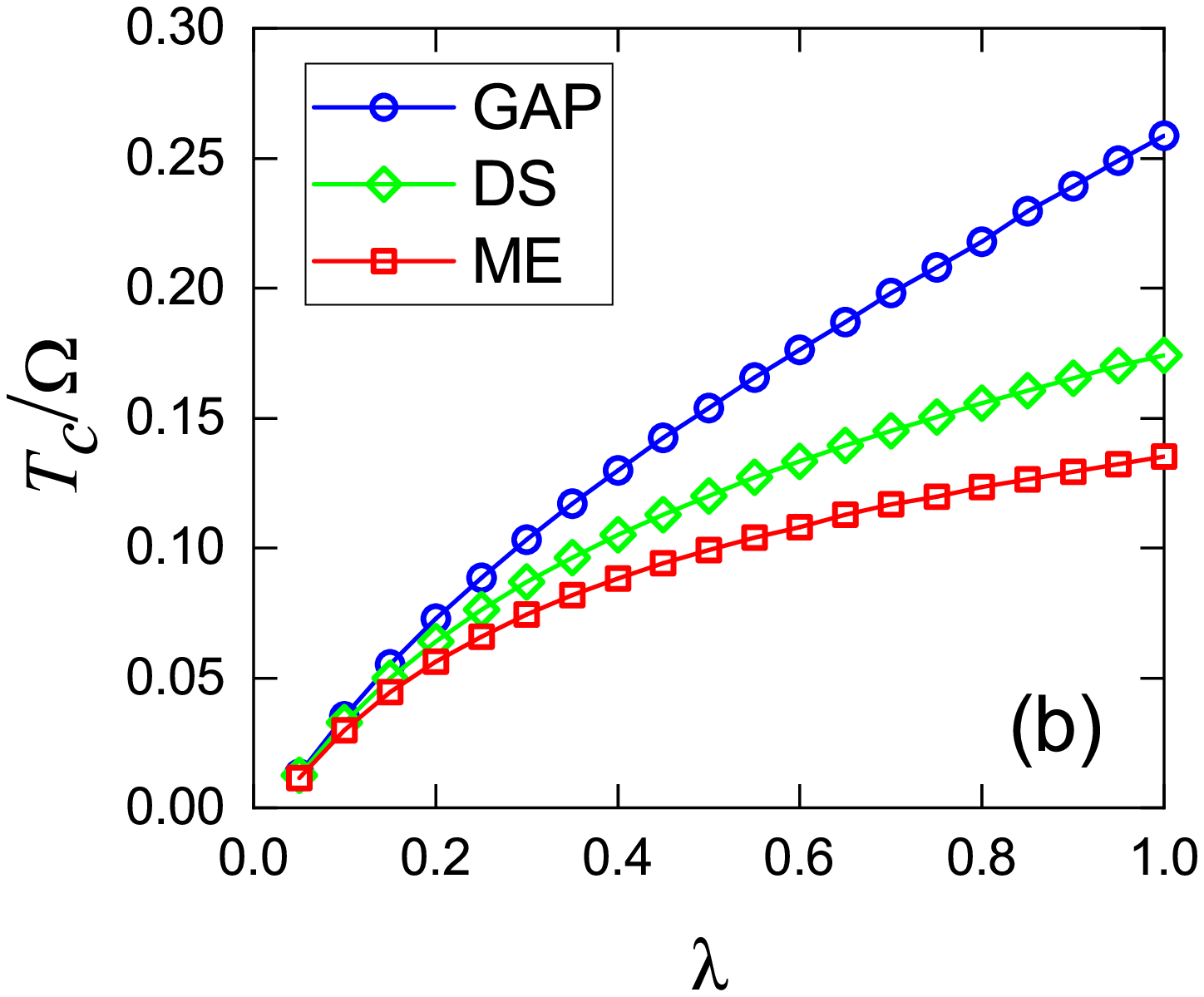}
\caption{Numerical results of $T_c$. (a): using $\delta$-function of
$g(\mathbf{q})$ (corresponding to $a_{0} \to 0$ limit). (b): using
exponential function of $g(\mathbf{q})$. There is only a minor
difference between (a) and (b). Dimensionless coupling parameter
$\lambda$ is related to $g_{0}$ by the formula $g_{0} =
\sqrt{\lambda}\Omega$. If $\Omega = 81\mathrm{meV}$, $0.01\Omega
\approx 9.5\mathrm{K}$.} \label{numericaltc}
\end{figure}

Motivated by the existing experiments, the IOP-induced
superconductivity has been theoretically investigated \cite{Xiang12,
Xing14, Johnston-NJP2016, Dolgov17, Opp-PRB2018} by solving the ME
equations of $A_{1}(\epsilon,\mathbf{p})$ and
$\Delta(\epsilon,\mathbf{p})$. However, to date, there is still no
consensus on the value of $T_c$ caused by IOPs. Different results
have been obtained by different groups. The existence of controversy
about $T_c$ is not surprising, as the kinetics and dynamics of
mobile electrons in FeSe film is quite complicated. In order to
acquire precise result of $T_c$, one needs to take into account
several effects, including the potentially important contributions
of vertex corrections, the multi-band electronic structure
\cite{Opp-PRB2018}, the unusual screening of Coulomb interaction
\cite{Millis18}, and the influence of magnetic and nematic
fluctuations \cite{Yao16}. Among all of these effects, the role of
vertex corrections is of special interest. Experiments \cite{Shen14}
and first principles calculations \cite{Gongxg15} have confirmed
that IOPs are nearly dispersionless, and the frequency
$\Omega(\mathbf{q})\equiv \Omega \approx 81 \mathrm{meV}$. In
comparison, the Fermi energy \cite{Lee15} is roughly $E_{F}\approx
65$meV. Apparently, the Migdal theorem is no longer reliable since
the ratio $\Omega/E_F > 1$, and the impact of vertex corrections
must be properly incorporated.

In this paper, we will not investigate all the effects mentioned
above. Our concentration is on the influence of vertex corrections
on the value of $T_c$. For simplicity, we consider the single band
model, following Ref.~\cite{Johnston-NJP2016}. It is known that the
scattering due to IOPs is dominated by small-$\mathbf{q}$ forward
scattering \cite{Shen14, Lee15, Johnston-NJP2016}, which is
described by a coupling function $g(\mathbf{q}) = g_0
e^{-|\mathbf{q}|/q_0}$ where the parameter $q_0$ is determined by
lattice constant. Similar to Ref.~\cite{Johnston-NJP2016}, we
consider two different forms of $g(\mathbf{q})$: $\delta$-function
and exponential function. The full set of integral equations of
$A_{1,2}(\epsilon,\mathbf{p})$ and $\Delta(\epsilon,\mathbf{p})$ are
too complicated to solve and proper approximations are unavoidable.
Since $A_{2}(\epsilon,\mathbf{p})$ can be absorbed into a
redefinition of chemical potential, here we assume that
$A_{2}(\epsilon,\mathbf{p})=1$. Because IOPs lead to extreme forward
scattering, the initial and final states of electrons are mostly
located near Fermi surface, which allows us to further set
$\xi_{\mathbf{p}}=\xi_{\textbf{p}_{F}}$. As a result, the coupled
equations are independent of momenta.

To separate the physical effects of the mass renormalization and the
vertex corrections, we will consider three approximations: (1) the
so-called GAP approximation that includes only the gap equation and
ignores both mass renormalization ($A_{1}(\epsilon)=1$) and vertex
corrections ($\Gamma_{t}=\sigma_{3}$); (2) the standard ME
approximation that couples the equation of $A_{1}(\epsilon)$
self-consistently to that of $\Delta(\epsilon)$ but ignores all the
vertex corrections ($\Gamma_{t}=\sigma_{3}$); (3) our DS equation
approximation that self-consistently takes into account the mass
renormalization $A_{1}(\epsilon)$ and all the vertex corrections.
The corresponding equations of $A_{1}(\epsilon)$ and
$\Delta(\epsilon)$ are explicitly shown in
Appendix~\ref{appendix:system}. These equations can be numerically
solved by using the iterative method, which is demonstrated in
Appendix~\ref{appendix:numerical}.

The numerical results of $T_c$ are shown in Fig.~\ref{numericaltc}.
The upper, middle, and lower curves are the results obtained under
GAP, DS, and ME approximations, respectively. We see that the GAP
approximation overestimates $T_c$, whereas the ME approximation
underestimates $T_c$. This indicates that, ignoring electron mass
renormalization may improperly increase $T_c$ and ignoring vertex
corrections may improperly reduce $T_c$. For very small values of
$\lambda$, GAP and ME results are both good. When $\lambda$ is
increasing, the deviation from DS results (middle curve) becomes
progressively more dramatic. For instance, if $\lambda=0.5$, the GAP
result of $T_c$ (denoted by $T_c^{\mathrm{GAP}}$) is about
$32\mathrm{K}$ higher, and the ME result of $T_c$ (denoted by
$T_c^{\mathrm{ME}}$) is about $20\mathrm{K}$ lower, than the DS
result of $T_c$ (denoted by $T_c^{\mathrm{DS}}$). For $\lambda=1.0$,
$T_c^{\mathrm{GAP}}-T_{c}^{\mathrm{DS}} \approx 85\mathrm{K}$ and
$T_c^{\mathrm{DS}}-T_{c}^{\mathrm{ME}}\approx 38\mathrm{K}$. For
$\lambda=2.0$ (not shown in Fig.~\ref{numericaltc}),
$T_c^{\mathrm{GAP}}-T_{c}^{\mathrm{DS}} \approx 200\mathrm{K}$ and
$T_c^{\mathrm{DS}}-T_{c}^{\mathrm{ME}}\approx 70\mathrm{K}$. It is
clear that the GAP and ME approximations are valid only in the
weak-coupling regime. When $\lambda$ becomes sufficiently large,
both GAP and ME approximations are quantitatively unreliable. In
contrast, our DS equation approach remains well-controlled and can
lead to reliable results.

The effect of Coulomb interaction is not included in the above
analysis. The Coulomb interaction is usually treated by the
pseudopotential method \cite{Allen, Carbotte}. After including the
contribution of pseudopotential $\mu^{\ast}$, the value of $T_c$
would be reduced \cite{Johnston-NJP2016} by roughly $10$\%-$20$\%.

It is known that a nematic order, which spontaneously breaks
$C_4$-symmetry down to $C_2$-symmetry, exists in bulk FeSe material
\cite{Coldea17, Bohmer17, Hoffman17}. Although the physical origin
of nematic order is still in fierce debate \cite{Fernandes14}, a
widely accepted notion is that the nematic order is induced by
electron-electron interaction, rather than lattice distortion. When
monolayer FeSe is placed on SrTiO$_3$ substrate, nematic order is
suppressed \cite{Hoffman17}, indicating that the interaction that
causes nematicity should not play an important role in 1UC
FeSe/SrTiO$_3$. But nematic fluctuation may not be negligible and
could help enhance superconductivity \cite{Yao16}. Similarly,
although there is no long-range magnetic order in bulk FeSe and 1UC
FeSe/SrTiO$_3$, the spin fluctuation may mediate Cooper pairing.
Recently, the cooperative effect of spin fluctuation and IOPs on
superconducting $T_c$ in 1UC FeSe/SrTiO$_3$ has been studied
\cite{Johnston21} using the ME formalism. It will be interesting to
investigate the coupling of electrons to nematic and/or spin
fluctuation by employing our DS equation approach.

The realistic electronic structure of FeSe film is more complicated
than the single band model considered in our work. The multi-band
effects \cite{Opp-PRB2018, Johnston21} cannot be ignored. The
impacts of nematic and spin fluctuations also need to be examined.
In this sense, our results do not provide a conclusive answer to the
underlying microscopic mechanism for the remarkable $T_c$
enhancement. But our approach does provide a firm basis for further
theoretical studies. Indeed, even if the multi-band effects and
additional pairing mechanism(s) are considered, the value of $T_c$
cannot be determined accurately if the full EPI vertex corrections
were not correctly taken into account. As shown in
Fig.~\ref{numericaltc}, ignoring vertex corrections leads to a
considerable underestimation of $T_c$. Once the full vertex function
is incorporated into the DS equations, one could proceed to consider
the multi-band effects as well as additional pairing mechanism(s).

In the future, we will generalize our approach from the simple
single-band model to more realistic multi-band models
\cite{Opp-PRB2018}. The EPI vertex function may become more complex
after considering the multi-band effects. The interplay of EPI and
Coulomb interaction also needs to be handled more carefully. These
issues will be addressed in forthcoming separate works.

\section{Summary and discussion}
\label{sec:SAD}

In summary, our DS equation approach provides a novel
nonperturbative framework for the theoretical study of
superconductivity induced by EPI of any strength. To testify its
efficiency, it would be interesting to apply it to a number of
well-defined models. As the simplest model describing EPI, Holstein
model \cite{Holstein59} have been extensively investigated
\cite{Scalettar89, Marsiglio90, Scalettar93, Johnston18} by various
methods (see Ref.~\cite{Johnston18} for a recent review). When EPI
becomes very strong, superconductivity is not the only instability
and EPI may lead to a CDW state \cite{Scalettar89, Marsiglio90,
Scalettar93, Johnston18}. An extended version of our approach can be
used to investigate the competition between superconductivity and
CDW in Holstein model. The results will be presented elsewhere.

The applicability of our approach is not restricted to EPI systems.
The DS equations and associated WTIs can be similarly constructed
and solved if phonon is replaced by other sorts of bosons. A
particularly interesting example is the U(1) gauge boson that
couples strongly to gapless fermions excited around the Fermi
surface of a two-dimensional strange metal \cite{Lee92, Nayak94,
Polchinski94, Altshuler94, Lee09}. In this case, the vertex
corrections would have more significant effects on physical
quantities than EPI systems, because the function
$A_{1}(\epsilon,\mathbf{p})$ exhibits singular behaviors in the
ultra-low energy region. The non-Fermi liquid behavior produced by
gauge interaction can be studied by solving the self-consistent
integral equation of $A_{1}(\epsilon,\mathbf{p})$ and
$A_{2}(\epsilon,\mathbf{p})$.

Heavy fermion system is another platform to apply our approach. The
role played by phonons in heavy fermion system was studied in
Ref.~\cite{Jarrell10} based on a periodic Anderson model combined
with Holstein model. The coherence temperature $T_{\mathrm{coh}}$ is
nearly unaffected by phonons if EPI is handled within ME
approximation \cite{Jarrell10}. But calculations using DFMT in
concert with a continuous-time QMC impurity solver found
\cite{Jarrell10} that the coupling of conduction electrons to
phonons leads to a strong reduction of $T_{\mathrm{coh}}$. This
indicates the complete failure of Migdal theorem in heavy fermion
system, since the ratio $T_{D}/T_{\mathrm{coh}}$, $T_{D}$ being the
temperature scale of Debye frequency, is no longer small. Actually,
even weak EPI suffices to cause Kondo breakdown \cite{Jarrell10}. It
would be of great interest to revisit this problem by using our
approach and examine how these results are affected by EPI vertex
corrections.

The interest of this paper is restricted to metals with a finite
Fermi surface. In Dirac semimetals that exhibit zero-dimensional
Fermi points, the low-energy fermionic excitations have more degrees
of freedom (such as spin, sublattice, and valley). There, the vertex
function should be determined by a larger number of coupled WTIs.
Our approach has recently been generalized \cite{Pan20} to deal with
the strong fermion-phonon interaction and long-range Coulomb
interaction in Dirac semimetals.

\section*{Acknowledgement}

G.-Z.L. thanks Q.-F. Chen, X. Li, C.-B. Luo, Y.-H. Wu, and H.-S.
Zong for helpful discussions. The work is supported by the National
Natural Science Foundation of China under Grants 11574285, 11674327,
U1532267, and U1832209.

G.-Z.L. motivated and designed the project and wrote the manuscript.
G.-Z.L., Z.-K.Y., and X.-Y.P. performed the analytical calculations
and analyzed the results. J.-R.W. performed the numerical
calculations.

\appendix

\section{Functional integral rules}\label{appendix:functional}

Our field-theoretic analysis is carried out within the framework of
functional integral. To help the readers understand the analysis, in
this appendix we list some basic rules of functional integration
that are used in our derivation. These rules are not new and can be
found in standard textbooks of quantum field theory \cite{Itzykson,
Peskin}.

All the correlation functions are generated from three important
quantities: the partition function $Z(\eta^{\dag},\eta,J)$, the
generating functional $W(\eta^{\dag},\eta,J)$, and the generating
functional $\Xi(\Psi,\Psi^{\dag},\phi)$. They are defined as
follows:
\begin{eqnarray}
Z(\eta^{\dag},\eta,J) &=& \int \mathcal{D}\phi\mathcal{D}
\Psi^{\dag}\mathcal{D}\Psi\exp\left(i\int\mathcal{L}(z)\right), \\
W(\eta^{\dag},\eta,J) &=& -i\ln Z(\eta^{\dag},\eta,J), \\
\Xi(\Psi,\Psi^{\dag},\phi) &=& W(\eta^{\dag},\eta,J) - \int
(\eta^{\dag}\Psi + \Psi^{\dag}\eta + J\phi).\nonumber \\
\end{eqnarray}
The following identities will be frequently used:
\begin{eqnarray}
&&\frac{\delta W}{\delta J} = \langle\phi\rangle, \quad \frac{\delta
W}{\delta \eta} = -\langle\Psi^\dag\rangle, \quad \frac{\delta
W}{\delta \eta^\dag}=\langle\Psi\rangle,\label{eq:Wbianfen} \\
&& \frac{\delta\Xi}{\delta\phi}=-J, \quad
\frac{\delta\Xi}{\delta\Psi}=\eta^\dag, \quad
\frac{\delta\Xi}{\delta\Psi^\dag}=-\eta.\label{eq:Gbianfen}
\end{eqnarray}
$W(\eta^{\dag},\eta,J)$ generates all the connected Green's
functions and $\Xi(\Psi,\Psi^{\dag},\phi)$ generates all the
irreducible proper vertices of electron-phonon coupling. For
instance, the full electron propagator $G(z-z^\prime)$ and full
phonon propagator $F(z-z^\prime)$ are given by
\begin{eqnarray}
G(z-z^\prime) &\equiv& -i\langle\Psi(z)\Psi^\dag(z^\prime)\rangle =
\frac{\delta^2W}{\delta\eta^\dag(z)\delta\eta(z^\prime)},
\label{eq:E-Propagator1} \\
F(z-z^\prime) &\equiv& -i\langle\phi(z)\phi^\dag(z^\prime)\rangle =
-\frac{\delta^2W}{\delta J(z)\delta J(z^\prime)}.
\end{eqnarray}
With the help of Eqs.~(\ref{eq:Wbianfen}-\ref{eq:Gbianfen}), the
above two expressions are calculated by the following steps:
\begin{eqnarray}
\frac{\delta^2W}{\delta\eta^\dag(z)\delta\eta(z^\prime)} &=& -
\frac{\delta\Psi^\dag(z^\prime)}{\delta\eta^\dag(z)} = -
\left(\frac{\delta\eta^\dag(z)}{\delta
\Psi^\dag(z^\prime)}\right)^{-1}
\nonumber \\
&=& -\left(\frac{\delta^2\Xi}{\delta\Psi^{\dag}(z^\prime)
\delta\Psi(z)}\right)^{-1}, \\
\frac{\delta^2W}{\delta J(z)\delta J(z^\prime)} &=&
\frac{\delta\phi(z^\prime)}{\delta J(z)} = \left(\frac{\delta
J(z)}{\delta\phi(z^\prime)}\right)^{-1}
\nonumber \\
&=& -\left(\frac{\delta^2\Xi}{\delta\phi(z^\prime)
\delta\phi(z)}\right)^{-1}.\label{eq:P-Propagator2}
\end{eqnarray}
Here it is important to emphasize that $\frac{\delta^2W}{\delta
\eta^{\dag}\delta \eta}$ and $\frac{\delta^2W}{\delta J\delta J}$
only involve connected Feynman diagrams for the electron and phonon
propagators. To understand this, we take the phonon field $\phi$ as
an example, and perform functional derivatives as follows:
\begin{eqnarray}
\frac{\delta^2W}{\delta J\delta J} &=& -i\frac{\delta}{\delta J}
\left(\frac1{Z}\int\mathcal{D}\phi\mathcal{D}\Psi\mathcal{D}
\Psi^\dag i\phi \exp\{i\mathcal{L}\}\right)\nonumber \\
&=& i\frac{\int\mathcal{D}\phi\mathcal{D}\Psi\mathcal{D}\Psi^\dag
\phi \phi\exp\{i\mathcal{L}\}}{Z} \nonumber \\
&& -i\left(\frac{\int\mathcal{D}\phi
\mathcal{D}\Psi\mathcal{D}\Psi^\dag \phi
\exp\{i\mathcal{L}\}}{Z}\right)^{2}.
\end{eqnarray}
In the last line, the first term contains all the connected and
disconnected diagrams, whereas the second term contains all the
disconnected diagrams. Hence, the phonon propagator $F =
-\frac{\delta^2W}{\delta J\delta J}$ contains only connected
diagrams. The same is true for the electron propagator, namely $G =
\frac{\delta^2W}{\delta \eta^{\dag}\delta \eta}$ contains only
connected diagrams.

\begin{figure}[htbp]
\centering
\includegraphics[width=2.1in]{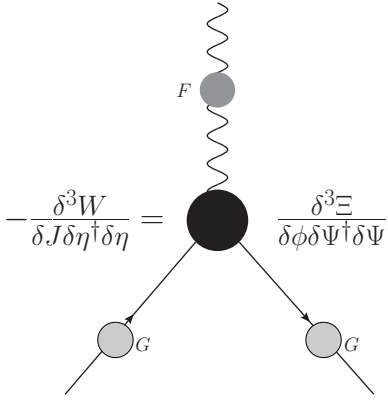}
\caption{Diagrammatic illustration of the relation given by
Eq.~(\ref{eq:EPIvertex}). Dropping three external legs of $3$-point
correlation function $\langle\phi\Psi\Psi^\dag\rangle$ yields the
EPI vertex function $\Gamma_{\mathrm{v}}$.} \label{vertexdiagram}
\end{figure}

The correlation function $\langle\phi\Psi\Psi^\dag\rangle$ is
expressed as follows
\begin{eqnarray}
\langle\phi \Psi\Psi^\dag\rangle \equiv \frac{\delta^3W}{\delta
J\delta\eta^\dag\delta\eta} = -F G \frac{\delta^3\Xi} {\delta\phi
\delta\Psi^\dag \delta\Psi}G,\label{eq:EPIvertexapp}
\end{eqnarray}
where we have defined a truncated (external legs being removed) EPI
vertex function $\Gamma_{\mathrm{v}}$, i.e.,
\begin{eqnarray}
\Gamma_{\mathrm{v}}(z_1,z_2,z_3) \equiv
\frac{\delta^3\Xi}{\delta\phi(z_1)\delta
\Psi^\dag(z_2)\delta\Psi(z_3)}.
\end{eqnarray}
These two equations are schematically illustrated by
Fig.~(\ref{vertexdiagram}). Analogous to the electron and phonon
propagators, here the vertex function $\Gamma_{\mathrm{v}}$ receives
contributions solely from connected diagrams. It is straightforward
to define and analyze four-point and higher-point correlation
functions by means of similar operations. However, for our purposes
it suffices to consider two- and three-point correlation functions.
For a more comprehensive illustration of functional integral
techniques, we would refer readers to the textbook of Itzykson and
Zuber \cite{Itzykson}.

\section{Dyson-Schwinger equations of fermion and boson propagators}
\label{appendix:DS}

In this appendix, we derive the formal DS integral equations of the
full fermion propagator $G(p)$ and the full phonon propagator $F(q)$
by using the rules of functional integral presented in
Appendix~\ref{appendix:functional}.

First derive the DS equation of electron propagator. The derivation
is based on an apparent fact that the partition function is
invariant under an arbitrary infinitesimal variation of spinor field
$\Psi^{\dag}$, that is
$$\int \mathcal{D}\phi\mathcal{D}\Psi^{\dag}\mathcal{D}\Psi
\frac{\delta}{i\delta\Psi^\dag}\exp\{iS\} = 0.$$ It is easy to get
\begin{eqnarray}
\langle(i\sigma_0\partial_{t}-\sigma_3\xi_{\partial_{\mathbf{z}}})
\Psi(z) - g \phi(z) \sigma_3 \Psi(z) +\eta(z)\rangle = 0.
\end{eqnarray}
Using the relations given by Eq.~(\ref{eq:Wbianfen}), we re-write
the above equation as
\begin{widetext}
\begin{eqnarray}
-\eta(z) = (i\sigma_0\partial_{t}
-\sigma_3\xi_{\partial_{\mathbf{z}}}) \frac{\delta W}
{\delta\eta^\dag(z)} +i g\sigma_3 \frac{\delta^2W}{\delta J(z)
\delta \eta^\dag(z)}-g\frac{\delta W}{\delta J(z)}\sigma_3
\frac{\delta W}{\delta\eta^\dag(z)}.
\end{eqnarray}
The last term of r.h.s. vanishes upon removing sources and can be
directly omitted. Operating functional derivative of both sides with
respect to $\eta(z_2)$ yields
\begin{eqnarray}
\delta(z-z_2) \sigma_0 &=& (i\sigma_0\partial_{t} - \sigma_3
\xi_{\partial_{\mathbf{z}}})\frac{\delta^2 W}{\delta\eta^\dag(z)
\delta\eta(z_2)}+i g \sigma_3 \frac{\delta^3W}{\delta J(z)
\delta\eta^\dag(z)\delta\eta(z_2)} \nonumber \\
&=& (i\sigma_0\partial_{t} -\sigma_3 \xi_{\partial_{\mathbf{z}}})
G(z-z_2) - ig\int dz_1^{\prime} d z_2^{\prime}\sigma_3
F(z-z_1^{\prime})G(z-z^{\prime})\frac{\delta^3\Xi}{\delta
\phi(z_1^\prime)\delta\Psi^\dag(z^\prime)\delta\Psi(z_2^\prime)}
G(z_2^\prime-z_2).\nonumber
\end{eqnarray}
This expression can be re-written as
\begin{eqnarray}
G^{-1}(z-z_3) &=& (i\sigma_0\partial_{t} - \sigma_3
\xi_{\partial_{\mathbf{z}}})\delta(z-z_3) - ig\int dz_1^\prime
dz^\prime\sigma_3 F(z-z_1^\prime)G(z-z^\prime) \frac{\delta^3\Xi}
{\delta\phi(z_1^\prime)\delta\Psi^\dag(z^\prime)\delta\Psi(z_3)}.
\label{Eq:DSEGPZ}
\end{eqnarray}
\end{widetext}
Making use of the definition
\begin{eqnarray}
\Gamma_{\mathrm{v}}(z_1^{\prime},z^{\prime},z_3) \equiv
\frac{\delta^3\Xi}{\delta\phi(z_1^{\prime})\delta
\Psi^\dag(z^\prime)\delta\Psi(z_3)}
\end{eqnarray}
and performing Fourier transformation to Eq.~(\ref{Eq:DSEGPZ}), we
ultimately obtain the following DS equation for fermion propagator
$G(p)$:
\begin{eqnarray}
G^{-1}(p)=G_{0}^{-1}(p)-i g\int dq\sigma_3G(p+q)F(q)
\Gamma_{\mathrm{v}}(q,p).
\end{eqnarray}
The Fourier transformation of
$\Gamma_{\mathrm{v}}(z_1^{\prime},z^{\prime},z_3)$ is
\begin{eqnarray}
\Gamma_{\mathrm{v}}(q,p) = \int dz_1^\prime dz^\prime
e^{ip(z_1^\prime-z_3)}e^{-i(p+q)(z_1^\prime-z^\prime)}
\Gamma_{\mathrm{v}}(z_1^\prime,z^\prime,z_3).\nonumber
\label{eq:gammavfourier}
\end{eqnarray}

\begin{figure}[htbp]
\centering
\includegraphics[width=3.39in]{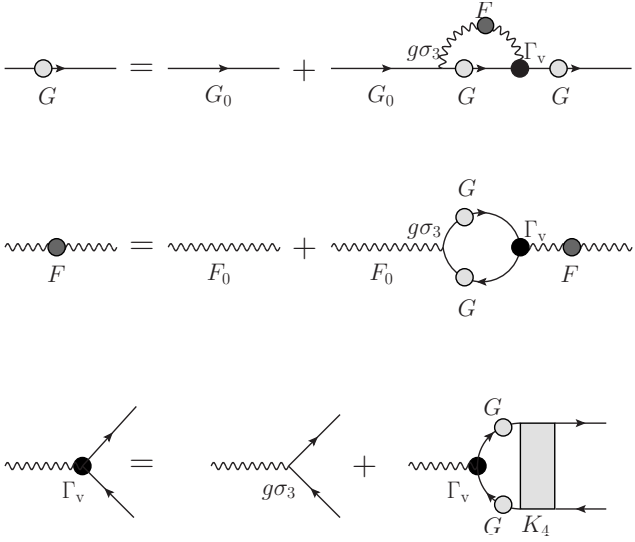}
\caption{An intuitive diagrammatic illustration of the coupled DS
equations of $G(p)$, $F(q)$, and $\Gamma_{\mathrm{v}}(q,p)$. The
complicated relation of these three functions can be readily seen.
The connection between $K_{4}(p,p',q)$ and higher-point correlation
functions are not shown.} \label{DSESdiagrams}
\end{figure}

The DS integral equation of $F(q)$ is derived based on the fact that
the partition function is invariant under an arbitrary infinitesimal
variation of phonon field $\phi$, i.e.,
$$\int \mathcal{D}\phi\mathcal{D}\Psi^{\dag}\mathcal{D}\Psi
\frac{\delta}{i\delta\phi}\exp\{iS\} = 0.$$ Then one finds that
\begin{eqnarray}
\langle \mathbb{D}_{z}\phi(z) -
g\Psi^\dag(z)\sigma_3\Psi(z)+J(z)\rangle = 0.
\end{eqnarray}

Again, we use the relations of Eq.~(\ref{eq:Wbianfen}) to obtain
\begin{eqnarray}
-J(z) &=& \mathbb{D}_{z}\langle\phi(z)\rangle +
ig\mathrm{Tr}\left[\sigma_3 \frac{\delta^2W}{\delta\eta^\dag(z)
\delta\eta(z)}\right] \nonumber \\
&& -g\mathrm{Tr}\left[\sigma_3 \frac{\delta W}{\delta
\eta^\dag(z)}\frac{\delta W}{\delta\eta(z)}\right].
\end{eqnarray}
Here a trace operator is introduced to operate on the components of
Nambu spinors. The last term of r.h.s. vanishes upon removing
external sources and can be directly omitted. This expression can be
re-written, using rules of functional integral, as
\begin{eqnarray}
\frac{\delta\Xi}{\delta\phi(z)} =
\mathbb{D}_{z}\langle\phi(z)\rangle - ig\mathrm{Tr}\left[\sigma_3
\left(\frac{\delta^2\Xi}{\delta\Psi^{\dag}(z)
\delta\Psi(z)}\right)^{-1}\right].\nonumber
\end{eqnarray}
Performing functional derivative of both sides of this equation with
respect to the field $\phi(z^\prime)$ gives rise to
\begin{widetext}
\begin{eqnarray}
\frac{\delta^2\Xi}{\delta\phi(z)\delta\phi(z^\prime)} =
\mathbb{D}_{z}\delta(z-z^\prime) +ig\int dz_1 dz_2
\mathrm{Tr}\left[\sigma_3 \left(\frac{\delta^2 \Xi}{\delta
\Psi^{\dag}(z_1)\delta\Psi(z)}\right)^{-1}
\left(\frac{\delta^3\Xi}{\delta\phi(z^\prime)\delta\Psi^{\dag}(z_1)
\delta\Psi(z_2)}\right)\left(\frac{\delta^2
\Xi}{\delta\Psi^{\dag}(z)\delta\Psi(z_2)}\right)^{-1}\right].
\nonumber
\end{eqnarray}
Making use of the identities given by
Eqs.~(\ref{eq:E-Propagator1}-\ref{eq:P-Propagator2}), we obtain
\begin{eqnarray}
F^{-1}(z-z^\prime) = \mathbb{D}_{z}\delta(z-z^\prime) +ig\int dz_1
dz_2 \mathrm{Tr}\left[\sigma_3 G(z-z_1) \Gamma(z^\prime,z_1,z_2)
G(z_2 - z)\right],
\end{eqnarray}
which, after carrying our Fourier transformation according to
Eq.~(\ref{eq:gammavfourier}), becomes
\begin{eqnarray}
F^{-1}(q) = F_{0}^{-1}(q) + i g\int_{p}\mathrm{Tr}\left[\sigma_3
G(p+q)\Gamma_{\mathrm{v}}(q,p)G(p)\right].\nonumber \\
\label{eq:originaldsphonon}
\end{eqnarray}
This is the DS equation of the full boson propagator.

The DS equations of the EPI vertex function
$\Gamma_{\mathrm{v}}(q,p)$ can be derived by performing a series of
analogous calculations \cite{Itzykson}. The derivation is quite
lengthy and the details will not be explicitly presented here. For a
more understandable diagrammatic illustration of the coupled DS
equations of $G(p)$, $F(q)$, and $\Gamma_{\mathrm{v}}(q,p)$, please
see Fig.~\ref{DSESdiagrams}. The DS equation of
$\Gamma_{\mathrm{v}}(q,p)$ can be formally written as
\begin{eqnarray}
\Gamma_{\mathrm{v}}(q,p) &=& g\sigma_3 - \int_{p'}G(p'+q)
\Gamma_\mathrm{v}(q,p')G(p')K_{4}(p,p',q).\label{eq:dsepivertex}
\nonumber \\
\end{eqnarray}
Here, the function $K_{4}(p,p',q)$ is defined via the $4$-point
correlation function $\langle \phi \phi^{\dag}\Psi
\Psi^{\dag}\rangle$ as follows: $$FF K_4 GG = \langle \phi
\phi^{\dag}\Psi \Psi^{\dag}\rangle.$$ One could verify that, the
function $K_{4}(p,p',q)$ satisfies its own DS integral equation,
which in turn is related to other higher-point correlation
functions. In fact, there exists an infinite number of DS equations
that connect every $n$-point correlation function to a $(n+1)$-point
correlation function for all positive integers $n>1$. All of these
DS equations are self-consistently coupled. Therefore, the full set
of DS equations are not closed and cannot be tackled. For this
reason, although the full set of DS equations are exact and, in
principle, contain all the interaction-induced effects, they are
rarely used in the realistic studies on strongly correlated electron
systems. Fortunately, we have shown in the main paper that the DS
equation of electron propagator $G(p)$ is indeed self-closed if
several symmetry-induced exact identities are properly taken into
account. The full boson propagator $F(q)$ and the vertex function
$\Gamma_{\mathrm{v}}(q,p)$ can be obtained from the numerical
solutions of $G(p)$.

\section{Gap equations in small-$\mathbf{q}$ dominated EPI systems}
\label{appendix:system}

In 1UC FeSe/SrTiO$_3$, electrons in FeSe film couple to IOPs.
Different from ordinary (acoustic) phonons, IOPs are nearly
dispersionless. The dispersion $\Omega_{\mathbf{q}}$ can thus be
taken as a constant. This type of EPI is sharply peaked at
$\mathbf{q}=0$ \cite{Shen14, Lee15, Johnston-NJP2016, Dolgov17,
Opp-PRB2018}. We will make use of this unique feature to simplify
the vertex function $\Gamma_{t}(q,p)$, which then reduces the time
required to complete the numerical computation. Moreover, if one is
mainly interested in the determination of $T_c$, it is reasonable to
linearize the DS equations, i.e., taking the $\Delta \to 0$ limit,
near $T_c$.

As discussed in Sec.~\ref{sec:application}, here we will consider
three different approximations, namely GAP, ME, and DS
approximations, of the coupled integral functions of $A_1(\epsilon)$
and $\Delta(\epsilon)$. To compare to results reported in
Ref.~\cite{Johnston-NJP2016}, we also consider two different forms
of coupling function $g(\mathbf{q})$: an idealized $\delta$-function
and a more realistic exponential function. We adopt Matsubara
formalism to describe finite-temperature correlation functions. The
electron frequency is $\epsilon_{n} = (2n+1)\pi T$ and the phonon
frequency is $\omega_{m} = 2m\pi T$, where $n$ and $m$ are integers.

\subsection{$\delta$-function}

In the case of $\delta$-function, the coupling function has the
simple form $g(\mathbf{q}) = g_{0}\delta(\mathbf{q})$. Under GAP
approximation, we take $A_{1}=1$. Then there is only the equation of
pairing function:
\begin{eqnarray}
\Delta(\epsilon_{n}) = \lambda\Omega^{2}T\sum_{m}
\frac{2\Omega}{\Omega^{2}+\omega_{m}^{2}}
\frac{\Delta(\epsilon_{n}+\omega_{m})}{(\epsilon_{n}+\omega_{m})^2
+\Delta^{2}(\epsilon_{n}+\omega_{m})}.\nonumber \\
\label{eq:GAPequation}
\end{eqnarray}
The dimensionless coupling constant $\lambda$ is related to EPI
coupling constant $g_{0}$ by $g_{0} = \sqrt{\lambda}\Omega$. This
approximation ignores the mass renormalization and supposes that EPI
leads only to Cooper pairing. The gap equation given by
Eq.~(\ref{eq:GAPequation}) is similar, but not identical to, the
standard BCS gap equation. This equation has previously been
analyzed by Rademaker \emph{et al.} \cite{Johnston-NJP2016}, who
made a comparison between the solution of Eq.~(\ref{eq:GAPequation})
to that of the standard BCS gap equation and shown that extreme
forward scattering leads to a remarkable enhancement of $T_c$.

Certainly, it is inappropriate to neglect the contributions of
$A_{1}$. If both $A_{1}$ and $\Delta$ are considered, we would
obtain the following two coupled ME equations:
\begin{eqnarray}
\Delta(\epsilon_{n}) &=& \lambda\Omega^{2}T\sum_{m}
\frac{2\Omega}{\Omega^{2}+\omega_{m}^{2}}
\frac{\Delta(\epsilon_{n}+\omega_{m})}{A_{1}^2(\epsilon_{n} +
\omega_{m})(\epsilon_{n}+\omega_{m})^2 +
\Delta^{2}(\epsilon_{n}+\omega_{m})}, \\
A_{1}(\epsilon_{n}) &=& 1+\frac{1}{\epsilon_{n}}
\lambda\Omega^{2}T\sum_{m} \frac{2\Omega}{\Omega^{2} +
\omega_{m}^{2}} \frac{\epsilon_n+\omega_{m}}{A_{1}^2
(\epsilon_{n}+\omega_{m})(\epsilon_{n}+\omega_{m})^2 +
\Delta^{2}(\epsilon_{n}+\omega_{m})}.
\end{eqnarray}
Including the full vertex corrections, described by $\Gamma_{t}$, to
the above ME equations leads to the following two DS equations
\begin{eqnarray}
\Delta(\epsilon_{n}) &=& \lambda\Omega^{2}T\sum_{m}
\frac{2\Omega}{\Omega^{2}+\omega_{m}^{2}}\frac{\Delta(\epsilon_{n} +
\omega_{m})}{A_{1}^2(\epsilon_{n}+\omega_{m})(\epsilon_{n}+\omega_{m})^2
+ \Delta^{2}(\epsilon_{n}+\omega_{m})}
\frac{A_{1}(\epsilon_{n}+\omega_{m})(\epsilon_n +
\omega_{m})-A_{1}(\epsilon_{n})\epsilon_{n}}{\omega_{m}}, \\
A_{1}(\epsilon_{n}) &=& 1+\frac{1}{\epsilon_{n}} \lambda
\Omega^{2}T\sum_{m}\frac{2\Omega}{\Omega^{2}+\omega_{m}^{2}}
\frac{A_{1}(\epsilon_{n}+\omega_{m})(\epsilon_n+\omega_{m})}{A_{1}^2
(\epsilon_{n}+\omega_{m})(\epsilon_{n}+\omega_{m})^2 +
\Delta^{2}(\epsilon_{n}+\omega_{m})}
\frac{A_{1}(\epsilon_{n}+\omega_{m})(\epsilon_n+\omega_{m}) -
A_{1}(\epsilon_{n})\epsilon_{n}} {\omega_{m}}.\nonumber \\
\end{eqnarray}

\subsection{Exponential function}

We then consider the more realistic exponential function of coupling
parameter $g(\mathbf{q}) = \sqrt{\lambda}\Omega
e^{-\mathbf{q}/q_0}$. It is necessary to introduce an UV cutoff
$\Lambda$, which then can be used to define a dimensionless
parameter $r = q_0/\Lambda$. Under the GAP approximation, the pure
gap equation is given by
\begin{eqnarray}
\Delta(\epsilon_{n}) &=& \left(\frac{2}{r}\right)^{2}
\lambda\Omega^{2}T\sum_{m}\int_{0}^{1}dxx \exp\left(-\frac{2
x}{r}\right)\frac{2\Omega}{\Omega^{2}+\omega_{m}^{2}}\nonumber
\\
&&\times\frac{\Delta(\epsilon_{n}+\omega_{m})}{\sqrt{(\epsilon_{n} +
\omega_{m})^2 + \Delta^{2}(\epsilon_{n}+\omega_{m})}
\sqrt{(\epsilon_{n}+\omega_{m})^2 + \Delta^{2}(\epsilon_{n} +
\omega_{m})+\zeta x^{2}}}.
\end{eqnarray}
The coupled ME equations of $\Delta(\epsilon_{n})$ and
$A_{1}(\epsilon_{n})$ are
\begin{eqnarray}
\Delta(\epsilon_{n}) &=& \left(\frac{2}{r}\right)^{2}
\lambda\Omega^{2}T\sum_{m} \int_{0}^{1}dx x
\exp\left(-\frac{2x}{r}\right)
\frac{2\Omega}{\Omega^{2}+\omega_{m}^{2}} \nonumber
\\
&&\times\frac{\Delta(\epsilon_{n}+\omega_{m})}
{\sqrt{A_1^2(\epsilon_{n}+\omega_{m})(\epsilon_{n} +
\omega_{m})^2 + \Delta^{2}(\epsilon_{n}+\omega_{m})}
\sqrt{A_1^2(\epsilon_{n}+\omega_{m})(\epsilon_{n}+\omega_{m})^2
+ \Delta^{2}(\epsilon_{n} + \omega_{m})+\zeta x^{2}}}, \\
A_{1}(\epsilon_{n}) &=& 1+\frac{1}{\epsilon_{n}}
\left(\frac{2}{r}\right)^{2}\lambda\Omega^{2}T\sum_{m}
\int_{0}^{1}dxx \exp\left(-\frac{2x}{r}\right)
\frac{2\Omega}{\Omega^{2}+\omega_{m}^{2}}\nonumber \\
&&\times\frac{A_{1}(\epsilon_{n}+\omega_{m})(\epsilon_n +
\omega_{m})}{\sqrt{A_1^2(\epsilon_{n}+\omega_{m})
(\epsilon_{n}+\omega_{m})^2 + \Delta^{2}(\epsilon_{n}+\omega_{m})}
\sqrt{A_1^2(\epsilon_{n}+\omega_{m})(\epsilon_{n}+\omega_{m})^2
+\Delta^{2}(\epsilon_{n}+\omega_{m})+\zeta x^{2}}}.
\end{eqnarray}
After including vertex corrections, the coupled DS equations are of
the form
\begin{eqnarray}
\Delta(\epsilon_{n}) &=& \left(\frac{2}{r}\right)^{2}
\lambda\Omega^{2}T\sum_{m}\int_{0}^{1}dx x
\exp\left(-\frac{2x}{r}\right)\frac{2\Omega}{\Omega^{2}+\omega_{m}^{2}}
\nonumber \\
&& \times \frac{\Delta(\epsilon_{n}+\omega_{m})}{\sqrt{
A_1^2(\epsilon_{n}+\omega_{m})(\epsilon_{n}+ \omega_{m})^2 +
\Delta^{2}(\epsilon_{n}+\omega_{m})}
\sqrt{A_1^2(\epsilon_{n}+\omega_{m})(\epsilon_{n}+\omega_{m})^2
+ \Delta^{2}(\epsilon_{n}+\omega_{m})+\zeta x^{2}}} \nonumber \\
&& \times \frac{A_{1}(\epsilon_{n} + \omega_{m})(\epsilon_n +
\omega_{m})-A_{1}(\epsilon_{n}) \epsilon_{n}}{\omega_{m}},
\label{eq:example1}
\\
A_{1}(\epsilon_{n}) &=& 1+\frac{1}{\epsilon_{n}}
\left(\frac{2}{r}\right)^{2}\lambda\Omega^{2}T\sum_{m}
\int_{0}^{1}dxx \exp\left(-\frac{2 x}{r}\right)
\frac{2\Omega}{\Omega^{2}+\omega_{m}^{2}} \nonumber
\\
&&\times\frac{A_{1}(\epsilon_{n}+\omega_{m})(\epsilon_n +
\omega_{m})}{\sqrt{A_1^2(\epsilon_{n}+\omega_{m})
(\epsilon_{n}+\omega_{m})^2 +
\Delta^{2}(\epsilon_{n}+\omega_{m})}\sqrt{A_1^2(\epsilon_{n}+\omega_{m})
(\epsilon_{n}+\omega_{m})^2
+ \Delta^{2}(\epsilon_{n}+\omega_{m})+\zeta x^{2}}} \nonumber \\
&& \times \frac{A_{1}(\epsilon_{n} +
\omega_{m})(\epsilon_n+\omega_{m})-A_{1}(\epsilon_{n})
\epsilon_{n}}{\omega_{m}}.\label{eq:example2}
\end{eqnarray}

\section{Numerical method}\label{appendix:numerical}

The self-consistent integral function(s) can be solved numerically
by using the iterative method. Let us take
Eqs.~(\ref{eq:example1}-\ref{eq:example2}) as an example to
illustrate how the iterative method works. In the first step, one
assumes some initial values of $A_{1}$ and $\Delta$. In the second
step, substitute these initial values into
Eqs.~(\ref{eq:example1}-\ref{eq:example2}) to obtain a set of new
values of $A_{1}$ and $\Delta$, which are more or less different
from the initial values. In the third step, substitute the new
values again into Eqs.~(\ref{eq:example1}-\ref{eq:example2}) to
obtain another set of new values of $A_{1}$ and $\Delta$. Repeat
this manipulation many times until the input and output of $A_{1}$
and $\Delta$ no longer change. Such stable values of $A_{1}$ and
$\Delta$ are precisely the solutions that we need.

The two equations contain a summation over $\omega_{m}$ for all
possible values of $m$. In practice, it is not possible, nor
necessary, to sum to infinity. On generic physical grounds we know
that $A_1$ and $\Delta$ are positive even functions of frequency.
Smaller frequency gives rise to larger $A_1$ and $\Delta$. When
electron frequency is much larger than phonon frequency $\Omega$,
the contributions are negligible. We choose a large number $N=400$
and define $\epsilon_{n}$ as follows
\begin{eqnarray}
\epsilon_n=\pm(2n-1)\pi T, \quad n=1,2,...,N.
\end{eqnarray}
Introduce $\omega_m^{\prime}$ and define it as $\omega^{\prime}_{m}
= \epsilon_{n} + \omega_{m}$. Thus $\omega^{\prime}_{m}$ is
restricted to the same region as $\epsilon_n$, namely
\begin{eqnarray}
\omega^{\prime}_m = \pm(2m-1)\pi T, && m=1,2,...,N.
\end{eqnarray}
Then Eqs.~(C9-C10) can be expressed as
\begin{eqnarray}
\Delta(\epsilon_{n}) &=& \left(\frac{2}{r}\right)^{2}
\lambda\Omega^{2}T\sum_{m=1}^{N}\int_{0}^{1}dx x
\exp\left(-\frac{2x}{r}\right)\frac{2\Omega}
{\Omega^{2}+(\omega^\prime_m-\epsilon_n)^{2}}
\nonumber \\
&& \times \frac{\Delta(\omega^\prime_m)}{\sqrt{A_{1}^{2}(
\omega^{\prime}_{m}){\omega^\prime_m}^2 +
\Delta^{2}(\omega^\prime_m)}
\sqrt{A_1^2(\omega^\prime_m){\omega^\prime_m}^2 +
\Delta^{2}(\omega^\prime_m)+\zeta x^{2}}}
\frac{A_{1}(\omega^\prime_m)\omega^{\prime}_{m}-A_{1}(\epsilon_{n})
\epsilon_{n}}{\omega^\prime_m-\epsilon_n}, \\
A_{1}(\epsilon_{n}) &=& 1+\frac{1}{\epsilon_{n}}
\left(\frac{2}{r}\right)^{2}\lambda\Omega^{2}T\sum_{m=1}^N
\int_{0}^{1}dxx \exp\left(-\frac{2 x}{r}\right)
\frac{2\Omega}{\Omega^{2}+(\omega^\prime_m-\epsilon_n)^{2}}
\nonumber
\\
&&\times\frac{A_{1}(\omega^\prime_m)\omega^\prime_m}{\sqrt{A_1^2(
\omega^\prime_m){\omega^\prime_m}^2 + \Delta^{2}(\omega^\prime_m)}
\sqrt{A_1^2(\omega^\prime_m){\omega^\prime_m}^2 +
\Delta^{2}(\omega^\prime_m)+\zeta x^{2}}}
\frac{A_{1}(\omega^\prime_m) \omega^\prime_m-A_{1}(\epsilon_{n})
\epsilon_{n}}{\omega^\prime_m-\epsilon_n}.
\end{eqnarray}
Now choose two initial values for unknown functions $A_1$ and
$\Delta$: $A_1 = 1$ and $\Delta=10^{-3}$. The Gaussian quadrature is
used to integrate over variable $x$. After $i$ times of iteration,
we would obtain $A_{1,i}$ and $\Delta_i$, which are then substituted
into the above two equations to obtain $A_{1,i+1}$ and
$\Delta_{i+1}$. Repeat such calculations until the difference
between $i$-results and $(i+1)$-results vanishes. $A_{1,i+1}$ and
$\Delta_{i+1}$ are related to $A_{1,i}$ and $\Delta_{i}$ via the
following equations:
\begin{eqnarray}
\Delta_{i+1}(n) &=& \left(\frac{2}{r}\right)^{2}
\lambda\Omega^{2}T\sum_{m=1}^{N}\int_{0}^{1}dx x
\exp\left(-\frac{2x}{r}\right)\frac{2\Omega}
{\Omega^{2}+4\pi^2T^2(m-n)^{2}}
\nonumber \\
&& \times \frac{\Delta_i(m)}{\sqrt{A_{1,i}^2(m)(2m-1)^2\pi^2T^2 +
\Delta_i^{2}(m)}\sqrt{A_{1,i}^2(m)(2m-1)^2\pi^2T^2
+ \Delta_i^{2}(m)+\zeta x^{2}}} \nonumber \\
&& \times \frac{A_{1,i}(m)(2m-1)-A_{1,i}(n)(2n-1)}{2(m-n)}
\nonumber\\&& +\left(\frac{2}{r}\right)^{2} \lambda
\Omega^{2}T\sum_{m=1}^{N}\int_{0}^{1}dx x
\exp\left(-\frac{2x}{r}\right)
\frac{2\Omega}{\Omega^{2}+4\pi^2T^2(m+n)^{2}}
\nonumber \\
&& \times \frac{\Delta_i(m)}{\sqrt{A_{1,i}^2(m)(2m-1)^2\pi^2 T^2 +
\Delta_i^{2}(m)} \sqrt{A_{1,i}^2(m)(2m-1)^2\pi^2 T^2 +
\Delta_i^{2}(m)+\zeta x^{2}}} \nonumber \\
&& \times \frac{A_{1,i}(m)(2m-1)+A_{1,i}(n) (2n-1)}{2(m+n)},
\\
A_{1,i+1}(n) &=& 1+\frac{1}{(2n-1)} \left(\frac{2}{r}\right)^{2}
\lambda\Omega^{2}T\sum_{m=1}^N \int_{0}^{1}dxx \exp\left(-\frac{2
x}{r}\right) \frac{2\Omega}{\Omega^{2}+4\pi^2T^2(m-n)^{2}} \nonumber
\\
&&\times\frac{A_{1,i}(m)(2m-1)}
{\sqrt{A_{1,i}^2(m)(2m-1)^2\pi^2T^2 +
\Delta_i^{2}(m)}\sqrt{A_{1,i}^2(m)(2m-1)^2\pi^2T^2
+ \Delta_i^{2}(m)+\zeta x^{2}}} \nonumber \\
&& \times \frac{A_{1,i}(m)(2m-1)-A_{1,i}(n)(2n-1)}{2(m-n)} \nonumber
\\
&& -\frac{1}{(2n-1)}\left(\frac{2}{r}\right)^{2}\lambda \Omega^{2}
T\sum_{m=1}^N \int_{0}^{1}dxx \exp\left(-\frac{2 x}{r}\right)
\frac{2\Omega}{\Omega^{2}+4\pi^2T^2(m+n)^{2}} \nonumber
\\
&&\times\frac{A_{1,i}(m)(2m-1)}{\sqrt{A_{1,i}^2(m)(2m-1)^2\pi^2T^2 +
\Delta_i^{2}(m)}\sqrt{A_{1,i}^2(m)(2m-1)^2\pi^2T^2
+ \Delta_i^{2}(m)+\zeta x^{2}}} \nonumber \\
&& \times \frac{A_{1,i}(m)(2m-1)+A_{1,i}(n)(2n-1)}{2(m+n)}.
\end{eqnarray}

The error factors created after $i$ times of iteration are
\begin{eqnarray}
\mathrm{EPS}_{a}(i) = \frac{1}{N}\sum_{n=1}^{N}
\frac{|A_{1,i}(n)-A_{1,i-1}(n)|}{|A_{1,i}(n)| + |A_{1,i-1}(n)|},
\quad \mathrm{EPS}_{b}(i) = \frac{1}{N}\sum_{n=1}^{N}
\frac{|\Delta_{i}(n)-\Delta_{i-1}(n)|} {|\Delta_{i}(n)|+
|\Delta_{i-1}(n)|}.
\end{eqnarray}
For given values of $\lambda$ and $T$, both $\mathrm{EPS}_{a}(i)$
and $\mathrm{EPS}_{b}(i)$ decrease gradually with increasing $i$,
provided that $\Delta$ has nontrivial solutions. When
$\mathrm{EPS}_{a}(i)$ and $\mathrm{EPS}_{b}(i)$ become sufficiently
small, the iteration can be terminated and the finial results of
$A_{1}$ and $\Delta$ are obtained. In realistic calculations, we
take $\mathrm{EPS}_{a}<10^{-6}$ and $\mathrm{EPS}_{b}<10^{-6}$ as
the criterion for achieving convergence. If $\Delta$ does not have a
nontrivial solution, $\mathrm{EPS}_{a}$ still becomes gradually
small, but $\mathrm{EPS}_{b}$ does not tend to decrease with growing
$i$. Actually, $\Delta_{i}(n)$ would rapidly go to zero as $i$
grows. Once $\Delta_{i}(n)$, which takes a finite initial value,
becomes sufficiently small, we would take $\Delta(n)=0$ directly and
terminate the iterating process. In practice, the iteration
procedure can be terminated if
$\frac{1}{N}\sum_{n=1}^{N}\Delta_{i}(n)<10^{-9}$.
\end{widetext}


\begin{thebibliography}{99}


\bibitem{Schrieffer64}
J. R. Schrieffer, \emph{Theory of Superconductivity} (CRC Press,
2018).

\bibitem{AGD}
A. A. Abrikosov, L. P. Gor'kov, and I. Y. Dzyaloshinskii,
\emph{Quantum Field Theoretical Methods in Statistical Physics}
(Pergamon Press, 1965).

\bibitem{Scalapino}
D. J. Scalapino, \emph{The electron-phonon interaction and
strong-coupling superconductivity}, in \emph{Superconductivity},
edited by R. D. Parks (Marcel Dekker, New York, 1969).

\bibitem{Migdal}
A. Migdal, Interaction between electrons and lattice vibrations in a
normal metal, Sov. Phys. JETP {\bf 7}, 996 (1958).

\bibitem{Eliashberg}
G. M. Eliashberg, Interactions between electrons and lattice
vibrations in a superconductor, Sov. Phys. JETP {\bf 11}, 696
(1960).

\bibitem{Allen}
P. B. Allen and B. Mitrovi\'{c}, \emph{Theory of Superconducting
T$_{c}$}, Solid State Physics, Vol.37 (Academic Press, 1982).

\bibitem{Carbotte}
J. P. Carbotte, Properties of boson-exchange superconductors, Rev.
Mod. Phys. {\bf 62}, 1027 (1990).

\bibitem{Marsiglio19}
F. Marsiglio, Eliashberg theory: A short review, Ann. Phys. {\bf
417}, 168102 (2020).

\bibitem{Xue12}
Q.-Y. Wang, Z. Li, W.-H. Zhang, Z.-C. Zhang, J.-S. Zhang, W. Li, H.
Ding, Y.-B. Ou, P. Deng, K. Chang, J. Wen, C.-L. Song, K. He, J.-F.
Jia, S.-H. Ji, Y.-Y. Wang, L.-L. Wang, X. Chen, X.-C. Ma, and Q.-K.
Xue, Interface-induced high-temperature superconductivity in single
unit-cell FeSe films on SrTiO$_3$, Chin. Phys. Lett. {\bf 29},
037402 (2012).

\bibitem{Shen14}
J. J. Lee, F. T. Schmitt, R. G. Moore, S. Johnston, Y.-T. Cui, W.
Li, M. Yi, Z. K. Liu, M. Hashimoto, Y. Zhang, D. H. Lu, T. P.
Devereaux, D.-H. Lee, and Z.-X. Shen, Interfacial mode coupling as
the origin of the enhancement of $T_c$ in FeSe films on SrTiO$_3$,
Nature (London), {\bf 515}, 245 (2014).

\bibitem{Lee15}
D.-H. Lee, What makes the $T_c$ of FeSe/SrTiO3 so high? Chinese
Phys. {\bf 24}, 117405 (2015).

\bibitem{Gorkov16}
L. P. Gor'kov, Peculiarities of superconductivity in the
single-layer FeSe/SrTiO$_3$ interface, Phys. Rev. B {\bf 93},
060507(R) (2016).

\bibitem{Johnston-NJP2016}
L. Rademaker \emph{et al.}, Enhanced superconductivity due to
forward scattering in FeSe thin films on SrTiO$_3$ substrates, New
J. Phys. {\bf 18}, 022001 (2016).

\bibitem{Martin19}
I. Martin, Moir\'{e} superconductivity, Ann. Phys. {\bf 417}, 168118
(2020).

\bibitem{Engelsberg63}
S. Engelsberg and J. R. Schrieffer, Coupled Electron-Phonon System,
Phys. Rev. {\bf 131}, 993 (1963).

\bibitem{Alexandrov01}
A. S. Alexandrov, Breakdown of the Migdal-Eliashberg theory in the
strong-coupling adiabatic regime, Europhys. Lett. {\bf 56}, 92
(2001).

\bibitem{Kivelson18}
I. Esterlis, B. Nosarzewski, E. W. Huang, B. Moritz, T. P.
Devereaux, D. J. Scalapino, and S. A. Kivelson, Breakdown of the
Migdal-Eliashberg theory: A determinant quantum Monte Carlo study,
Phys. Rev. B {\bf 97}, 140501(R) (2018).

\bibitem{Schooley64}
J. F. Schooley, W. R. Hosler, and M. L. Cohen, Superconductivity in
semiconducting SrTiO$_3$, Phys. Rev. Lett. {\bf 12}, 474 (1964).

\bibitem{Chubukov19}
M. N. Gastiasoro, A. V. Chubukov, and R. M. Fernandes,
Phonon-mediated superconductivity in low carrier-density systems,
Phys. Rev. B {\bf 99}, 094524 (2019).

\bibitem{Caoyuan18}
Y. Cao, V. Fatemi, S. Fang, K. Watanabe, T. Taniguchi, E. Kaxiras
and P. Jarillo-Herrero, Unconventional superconductivity in
magic-angle graphene superlattices, Nature {\bf 556}, 43 (2018).

\bibitem{Gunnarsson97}
O. Gunnarsson, Superconductivity in fullerides, Rev. Mod. Phys. {\bf
69}, 575 (1997).

\bibitem{Cappelluti01}
E. Cappelluti, C. Grimaldi, L. Pietronero, S. Str\"{a}ssler, and
G.A. Ummarino, Superconductivity of Rb$_{3}$C$_{60}$: breakdown of
the Migdal-Eliashberg theory, Eur. Phys. J. B {\bf 21}, 383 (2001).

\bibitem{Shen05}
T. Cuk, D. H. Lu, X. J. Zhou, Z.-X. Shen, T. P. Devereaux, and N.
Nagaosa, A review of electron-phonon coupling seen in the high-Tc
superconductors by angle-resolved photoemission studies (ARPES),
phys. stat. sol. (b) {\bf 242}, 11 (2005).

\bibitem{Nambu60}
Y. Nambu, Quasi-particles and gauge invariance in the theory of
superconductivity, Phys. Rev. {\bf 117}, 648 (1960).

\bibitem{Itzykson}
C. Itzykson and J.-B. Zuber, \emph{Quantum Field Theory}
(McGraw-Hill, New York, 1980).

\bibitem{Takahashi86}
Y. Takahashi, in \emph{Quantum Field Theory}, edited by F. Mancini,
(Elsevier Science Publisher, 1986).

\bibitem{Kondo97}
K.-I. Kondo, Transverse Ward-Takahashi identity, anomaly, and
Schwinger-Dyson equation, Int. J. Mod. Phys. A, {\bf 12}, 5651
(1997).

\bibitem{He01}
H. He, F. C. Khanna, and Y. Takahashi, Transverse Ward-Takahashi
identity for the fermion-boson vertex in gauge theories, Phys. Lett.
B {\bf 480}, 222 (2000).

\bibitem{Dirac}
P. A. M. Dirac, Discussion of the infinite distribution of electrons
in the theory of the positron, Proc. Camb. Phil. Soc. {\bf 30}, 150
(1934).

\bibitem{Peierls}
R. Peierls, The vacuum in Dirac's theory of the positive electron,
Proc. Roy. Soc. Series A {\bf 146}, 420 (1934).

\bibitem{Serber}
R. Serber, A note on positron theory and proper energies, Phys. Rev.
{\bf 49}, 545 (1936).

\bibitem{Schwinger}
J. Schwinger, On gauge invariance and vacuum polarization, Phys.
Rev. {\bf 82}, 664 (1951).

\bibitem{Jackiw69}
R. Jackiw and K. Johnson, Anomalies of the axial-vector current,
Phys. Rev. {\bf 182}, 1459 (1969).

\bibitem{Callan70}
C. G. Callan, S. Coleman, and R. Jackiw, A new improved
energy-momentum tensor, Ann. Phys. {\bf 59}, 42 (1970).

\bibitem{Bardeen69}
W. A. Bardeen, Anonmalous Ward identities in spinor field theories,
Phys. Rev. {\bf 184}, 1848 (1969).

\bibitem{Peskin}
M. E. Peskin and D. V. Schroeder, \emph{An Introduction to Quantum
Field Theory} (CRC Press, 2019).

\bibitem{Schnabl}
J. Novotny and M. Schnabl, Point-splitting regularization of
composite operators and anomalies, Fortschr. Phys. {\bf 48}, 253
(2000).

\bibitem{Anderson63}
P. W. Anderson, Plasmons, gauge invariance, and mass, Phys. Rev.
{\bf 130}, 439 (1963).

\bibitem{Higgs64}
P. W. Higgs, Broken symmetries and the masses of gauge bosons, Phys.
Rev. Lett. {\bf 13}, 508 (1964).

\bibitem{Assaad08}
F. F. Assaad and H. G. Evertz, World-line and determinantal quantum
Monte Carlo methods for spins, phonons and electrons, Lect. Notes
Phys. {\bf 739}, 277 (2008).

\bibitem{DMFT}
A. Georges, G. Kotliar, W. Krauth, and M. J. Rozenberg, Dynamical
mean-field theory of strongly correlated fermion systems and the
limit of infinite dimensions, Rev. Mod. Phys. {\bf 68}, 13 (1996).

\bibitem{Wu08}
F.-C. Hsu, J.-Y. Luo, K.-W. Yeh, T.-K. Chen, T.-W. Huang, P. M. Wu,
Y.-C. Lee, Y.-L. Huang, Y.-Y. Chu, D.-C. Yan, and M.-K. Wu, From the
cover: Superconductivity in the PbO-type structure FeSe, Proc. Natl.
Acad. Sci. USA {\bf 105}, 14262 (2008).

\bibitem{Rebec17}
S. N. Rebec, T. Jia, C. Zhang, M. Hashimoto, D.-H. Lu, R. G. Moore,
and Z.-X. Shen, Coexistence of replica bands and superconductivity
in FeSe monolayer films, Phys. Rev. Lett. {\bf 118}, 067002 (2017).

\bibitem{Xiang12}
Y.-Y. Xiang, F. Wang, D. Wang, Q.-H. Wang, and D.-H. Lee,
High-temperature superconductivity at the FeSe/SrTiO$_3$ interface,
Phys. Rev. B {\bf 86}, 134508 (2012).

\bibitem{Xing14}
B. Li, Z. W. Xing, G. Q. Huang, and D. Y. Xing, Electron-phonon
coupling enhanced by the FeSe/SrTiO$_3$ interface, J. Appl. Phys.
{\bf 115}, 193907 (2014).

\bibitem{Dolgov17}
M. L. Kuli\'{c} and O. V. Dolgov, The electron-phonon interaction
with forward scattering peak is dominant in high $T_c$
superconductors of FeSe films on SrTiO$_{3}$ (TiO$_2$), New J. Phys.
{\bf 19}, 013020 (2017).

\bibitem{Opp-PRB2018}
A. Aperis and P. M. Oppeneer, Multiband full-bandwidth anisotropic
Eliashberg theory of interfacial electron-phonon coupling and
high-$T_c$ superconductivity in FeSe/SrTiO$_{3}$, Phys. Rev. B {\bf
97}, 060501(R) (2018).

\bibitem{Millis18}
Y. Zhou and A. J. Millis, Dipolar phonons and electronic screening
in monolayer FeSe on SrTiO$_3$, Phys. Rev. B {\bf 96}, 054516
(2017).

\bibitem{Yao16}
Z.-X. Li, F. Wang, D.-H. Lee, and H. Yao, What makes the $T_c$ of
monolayer FeSe on SrTiO$_3$ so high: a sign-problem-free quantum
Monte Carlo study, Sci. Bull. {\bf 61}, 925 (2016).

\bibitem{Gongxg15}
Y. Xie, H.-Y. Cao, Y. Zhou, S. Chen, H. Xiang, and X.-G. Gong,
Oxygen vacancy induced flat phonon mode at FeSe/SrTiO$_3$ interface,
Sci. Rep. {\bf 5}, 10011 (2015).

\bibitem{Coldea17}
A. Coldea and M. D. Watson, The Key Ingredients of the Electronic
Structure of FeSe, Ann. Rev. Condens. Matter Phys. {\bf 9}, 125
(2018).

\bibitem{Bohmer17}
A. E. B\"{o}hmer and A. Kreisel, Nematicity, magnetism and
superconductivity in FeSe, J. Phys. Condens. Matter {\bf 30}, 023001
(2017).

\bibitem{Hoffman17}
D. Huang and J. E. Hoffman, Monolayer FeSe on SrTiO$_{3}$, Annu.
Rev. Condens. Matter Phys. {\bf 8}, 311 (2017).

\bibitem{Fernandes14}
R. M. Fernandes, A. V. Chubukov, and J. Schmalian, \emph{What Drives
Nematic Order in Iron-Based Superconductors?}, Nat. Phys. {\bf 10},
97 (2014).

\bibitem{Johnston21}
L. Rademaker, G. Alvarez-Suchini, K. Nakatsukasa, Y. Wang, S.
Johnston, Enhanced superconductivity in FeSe/SrTiO$_3$ from the
combination of forward scattering phonons and spin fluctuations,
arXiv:2101.08307.

\bibitem{Holstein59}
T. Holstein, Studies of polaron motion: Part I. The
molecular-crystal model, Ann. Phys. {\bf 8}, 325 (1959).

\bibitem{Scalettar93}
P. Niyaz, J. E. Gubernatis, R. T. Scalettar, and C. Y. Fong,
Charge-density-wave-gap formation in the two-dimensional Holstein
model at half-filling, Phys. Rev. B {\bf 48}, 16011 (1993).

\bibitem{Scalettar89}
R. T. Scalettar, N. E. Bickers, and D. J. Scalapino, Competition of
pairing and Peierls-charge-density-wave correlations in a
two-dimensional electron-phonon model, Phys. Rev. B {\bf 40}, 197
(1989).

\bibitem{Marsiglio90}
F. Marsiglio, Pairing and charge-density-wave correlations in the
Holstein model at half-filling, Phys. Rev. B {\bf 42}, 2416 (1990).


\bibitem{Johnston18}
P. M. Dee, K. Nakatsukasa, Y. Wang, and S. Johnston,
Temperature-filling phase diagram of the two-dimensional Holstein
model in the thermodynamic limit by self-consistent Migdal
approximation, Phys. Rev. B {\bf 99}, 024514 (2018).

\bibitem{Lee92}
P. A. Lee and N. Nagaosa, Gauge theory of the normal state of
high-$T_c$ superconductors, Phys. Rev. B {\bf 46}, 5621 (1992).

\bibitem{Nayak94}
C. Nayak and F Wilczek, Non-Fermi liquid fixed point in $2+1$
dimensions, Nucl. Phys. B {\bf 417}, 359 (1994).

\bibitem{Polchinski94}
J. Polchinski, Low-energy dynamics of the spinon-gauge system, Nucl.
Phys. B {\bf 422}, 617 (1994).

\bibitem{Altshuler94}
B. L. Altshuler, L. B. Ioffe, and A. J. Millis, Low-energy
properties of fermions with singular interactions, Phys. Rev. B {\bf
50}, 14048 (1994).

\bibitem{Lee09}
S.-S. Lee, Low-energy effective theory of Fermi surface coupled with
U(1) gauge field in $2+1$ dimensions, Phys. Rev. B {\bf 80}, 165102
(2009).

\bibitem{Jarrell10}
M. Raczkowski, P. Zhang, F. F. Assaad, T. Pruschke, and M. Jarrell,
Phonons and the coherence scale of models of heavy fermions, Phys.
Rev. B {\bf 81}, 054444 (2010).

\bibitem{Pan20}
X.-Y. Pan, Z.-K. Yang, X. Li, and G.-Z. Liu, Nonperturbative
Dyson-Schwinger equation approach to strongly interacting Dirac
fermion systems, arXiv:2003.10371.


\end{thebibliography}
\end{document}